\documentclass[%
 reprint,
superscriptaddress,
 amsmath,amssymb,
 prl,
]{revtex4-2}
\usepackage{centernot}
\usepackage{graphicx}% Include figure files
\usepackage{dcolumn}% Align table columns on decimal point
\usepackage{bm}% bold math
\usepackage[dvipsnames]{xcolor}
\begin{document}

%\preprint{APS/123-QED}

\title{Double-replica theory for evolution of genotype-phenotype interrelationship}% Force line breaks with \\
%\thanks{A footnote to the article title}%

\author{Tuan Minh Pham}
\affiliation{The Niels Bohr Institute, University of Copenhagen,
Blegdamsvej 17, Copenhagen, 2100-DK, Denmark
}

\author{Kunihiko Kaneko}
\affiliation{The Niels Bohr Institute, University of Copenhagen,
Blegdamsvej 17, Copenhagen, 2100-DK, Denmark
}
\affiliation{Center for Complex Systems Biology, Universal Biology Institute,
University of Tokyo, Komaba, Tokyo 153-8902, Japan
}
\date{\today}% It is always \today, today,
             %  but any date may be explicitly specified

\begin{abstract}
The relationship between  genotype and phenotype plays a crucial role in determining the function and robustness of biological systems. Here the  evolution progresses through the change in genotype, whereas the selection is based on the phenotype, and  genotype-phenotype relation also evolves. Theory for such phenotypic evolution remains poorly-developed, in contrast to evolution under the fitness landscape determined by genotypes. Here we  provide statistical-physics formulation of this problem by introducing replicas for genotype and phenotype. We apply it to an evolution model, in which phenotypes are given by spin configurations; genotypes are interaction matrix for spins to give the Hamiltonian, and the  fitness depends only on the configuration of a subset of spins called target. We describe the interplay between the  genetic variations and phenotypic variances by noise in this model by our new approach that extends the replica theory for spin-glasses to include  spin-replica for phenotypes and coupling-replica for genotypes. Within this framework we obtain a phase diagram of the evolved phenotypes against the noise and selection pressure, where each phase is distinguished by the fitness and overlaps for genotypes and phenotypes. Among the phases, robust fitted phase, relevant to biological evolution,  is achieved under the intermediate level of noise (temperature), where robustness to noise and to genetic mutation are correlated, as a result of replica symmetry.  We also find a trade-off between maintaining a high fitness level of phenotype and acquiring a robust pattern of genes as well as the dependence of this trade-off on the ratio between the size of the functional (target) part to that of the remaining non-functional (non-target) one. The selection pressure needed to achieve  high fitness increases with the fraction of target spins.

\end{abstract}

%\keywords{Suggested keywords}%Use showkeys class option if keyword
                              %display desired
\maketitle

%\tableofcontents

\section{Introduction}
Over decades, evolution under given fitness landscape, determined as a function of genes (genotypes), has been studied extensively  \cite{Wright32, deVisser2014}. % Stadler2002
Here genotypes are changed (mutated) in the reproduction process, and those that produce higher function, called phenotypes, are selected. These phenotypes determine fitness, the rate of offsprings that survive. 
However, the evolution of phenotypes whose configurations are shaped by the genetic evolution remains poorly explored.  Here, phenotypes are a result
of dynamics whose rule is determined by genotypes. Such dynamics  are  stochastic in general. Cells involve stochastic gene expression dynamics, whereas protein folding dynamics to give the protein shape is under thermal noise \cite{McAdams, Elowitz, Furusawa2005, Bar-Even}. Phenotypes hence  are variable under noise, while fitted phenotypic states are better to be preserved under noise, i.e., to keep robustness to noise. In addition, they can also be varied by genetic mutation during the evolution of genotypes, and  the robustness to mutation will  also be required.  The achievement of robustness of phenotypes to noise and to mutation is important to the evolution, as has been discussed recently \cite{Wagner, Ancel,KanekoPloSOne2007, Shreif2014}. Now considering stochastic dynamics of phenotypes, a general formulation of the evolution of such genotype-phenotype mapping and phenotypic robustness is hence wanted.

Underlying such stochastic dynamics are the interactions among a vast number of elements that constitute a biological system.  A cell consists of a huge variety of interacting molecules and its constitute polymers (proteins) are composed of many monomers (residues). Now the states of such interacting elements that shape the evolution of phenotypes can be properly described  by statistical physics \cite{Toda, sethna2021statistical}. To this kind of study,  use of spin models  is  relevant, where  phenotypes are spin configurations that are updated by Hamiltonian with spin-spin interactions under thermal noise, whereas genotypes specify such spin-spin interactions, and fitness is given by a function of configuration of a subset of spins, termed as target spins. It is then important to identify possible phases with regards to genotypes and phenotypes, using the set of order parameters, a concept rooted in  statistical physics.

In the present paper, to  address these problems in a systematic way, we formulate double replica methods both for spins (phenotypes) and couplings (genotypes), as well as both for target and nontarget parts. Even though we adopted spin-coupling representation here  our formulation  can generally be applied to  other problems, in which the interactions among many degrees of freedom are also dynamical variables with their own dynamics. % the genotypes determine the interactions among many degrees of freedom, with which dynamics of phenotypes evolve to adapt to the external conditions.
For the sake of demonstration, however, we here  explain this approach by using specifically a spin-glass Hamiltonian model developed in \cite{Sakata2020,  Sakata2009, Sakata2009PRE}. These  works  uncovered the transitions between different regimes with regards to the fitness and robustness  upon changing the strength of thermal noise for phenotype and selection pressure for genotype. In particular, within an intermediate range of phenotypic noise, the evolved spin-systems could attain high fitness and robustness to noise and mutation. A systematic way to elucidate
the condition to achieve such robust fitted states with regards to the noise strength, selection pressure, and relative size of target spins, and to understand possible relationship between robustness to mutation and to noise, however, has not been developed yet. 
% We further need to understand the condition to achieve robust fitted states against the noise, selection pressure, and relative size of target spins, and possible relationship between the fitness and robustness, as well as that between robustness to mutation and to noise.  In this regard, it is desirable to have such an approach that could answer these questions in a systematic way.

One might  expect an application of  mean-field methods for disordered systems \cite{Mezard} in this model since a spin-glass Hamiltonian formulation was adopted  by replacing random couplings among spins (phenotypes) by genotypes that are  evolving \cite{Sakata2020,  Sakata2009, Sakata2009PRE}. Gradual change in the couplings might fit with partial annealing approach based on a finite number of replicas $n$. However,  the study is restricted to the case in which the coupling dynamics are affected by a spin-spin correlation term \cite{Penney1993,Coolen1993,Penney1994, Dotsenko1994, Uezu2009}, and is not directly applicable for our purpose, in which the coupling dynamics depend on  the fitness determined by the spin configurations.  Another theoretical method assumes the quench  limit ($n \rightarrow 0$) for  a replicated spin system % with   a division among replicas, due to fitness
%despite of genotypes  being dynamical, can provide a reasonable phase diagram of the model  
\cite{Sakata2012}. However, this means that the couplings are treated effectively as `static' (but with a modified distribution),  and hence is not suitable to investigate the evolution of both genotypes (couplings) and phenotypes (spin configurations). %only  the weights of different  spin configurations contribute to the system full partition function. 
Using this approach, it  thus  remains elusive to see how robustness emerges from the interplay between the ordering of spins and that of the couplings.

%THIS PART (LAST SENTENCE) IS TOO TECHNICAL FOR THE FIRST READER< SO AS A SENTENCE IN INTRODUCTION. IT WOULD NOT FIT. WE NEED TO EXPLAIN MORE SIMPLY WHY WE NEE THE THE PRESENT APPRAOACH INSTEAD OF THE PREVIOUS}

In this paper   we develop a new mean-field approach that we term as double replica theory. It describes the evolution of both genotypes and phenotypes by considering spins and links as two different replica species. With this formulation, we obtain fitness and replica overlaps for spins and couplings, which work as the order parameters. Using these order parameters we identify five regions in the temperature vs selection pressure phase diagram: two  non-fitted paramagnetic phases,  fitted- and non-fitted  spin glass phases,   and a robust
fitted phase.
%Interestingly, 
The last phase is the most biologically important, which can only be achieved under intermediate noise level (temperature) and sufficient selection pressure, whereas robustness can only arrive at the cost of lowering the fitness from its maximal  value.
%We quantify this effect in details as well as 
%making an account for 
Dependence of the robust fitted phase on the ratio of functional to non-functional parts has been analyzed in depth. As the former ratio is increased,  the selection pressure to achieve this phase  is drastically increased, whereas, if achieved, it can persist for slightly higher noise. This suggests the relevance of having sufficient non-functional parts in biological systems. 
In addition, correlation between robustness to noise and to mutation are formulated as a proportionality between susceptibilities to external field and to coupling change.
%LATER WE MAY NEED TO ADD A BIT, 

\section{Double replica theory}

Following \cite{Sakata2009, Sakata2009PRE} we study  the evolution of  the relationship between phenotype and genotype  by  representing phenotypes as spin configurations, and genotypes as interaction matrix for spins. In a system of $N$ spins,  each spin $i$ can take values $s_i \in \{-1,1\}$ and is linked to exactly $N-1$ other spins, thus forming a fully-connected network. Here the  evolution progresses through the change in genotype, whereas the selection is based on the phenotype, resulting in an evolution of the  genotype-phenotype relation. Moreover, fitness is determined by a subset of  target spins denoted by  $\mathcal{T}$.  Those spins that do not contribute to the fitness are called non-target. In general,  the fitness $\Psi$ is some field that acts on $J_{ij}$ but whose value depends only on $s_i$, $i\in \mathcal {T}$. How such dependence is explicitly described is model-specific and will not limit the use of our approach. 
%Assuming that fitness would be maximised if a global alignment is established among target spins, we introduce   
%\begin{equation}
%\Psi = \frac{1}{N_t}\, \Big|\sum_{i \in \mathcal{T}} s_i  \Big| \,, \label{fitness}
%  \end{equation}
%where $N_t$ is the size of $\mathcal{T}$. 
See SM Eq. \eqref{fitness_SM} for an example of $\Psi$  given by the target spin configurations at equilibrium  \cite{Sakata2009, Sakata2009PRE}.  

Stochastic dynamics of phenotypes are considered as  the  evolution of spin configurations at a temperature $T_s$ according to a Hamiltonian $H_S = -\sum_{i<j} J_{ij}s_i s_j$ \cite{Sherrington}. Here the couplings $J_{ij}$ are regarded as \emph{fixed} over the course of the spin evolution that follows a Glauber update because they  are assumed to evolve on much slower timescale than that of the spins. Furtheremore, the couplings are  symmetric, i.e.  $J_{ij} = J_{ji}$, and, \textit{initially}, are  independently and identically distributed by a Gaussian distribution with  zero mean and the variance $J^2 := {\rm var}(J_{ij}) = N^{-1}$. The coupling matrix  $\bold{J}$ includes interactions  between target spins  $J_{ij}^{(tt)}$ for $i \in \mathcal{T}$ and $j \in \mathcal{T}$;  those between non-target spins  $J_{ij}^{(oo)}$ for $i\centernot  \in \mathcal{T}$ and $j\centernot  \in \mathcal{T}$; and those between target spin  and non-target spin  $J_{ij}^{(to)}$ for $i \in \mathcal{T}$ and $j\centernot  \in \mathcal{T}$.  Let $S_{\mathcal{T}}$ and $S_{\mathcal{O}}$ denote the subsystem of target spins (with their interactions $\bold{J}^{(tt)}$) and the subsystem of   non-target spins (with the couplings $\bold{J}^{(oo)}$ among them), respectively. The  Hamiltonian of the full system denoted by $S$ can be decomposed into 
\begin{equation}H_{S} = \underbrace{- \sum_{i<j\in \mathcal{T}} J^{(tt)}_{ij} s_is_j}_{H_{\mathcal{T}}}  \underbrace{- \sum_{i<j \centernot \in \mathcal{T}} J^{(oo)}_{ij} s_is_j }_{  H_{\mathcal{O}}}  \underbrace{- \sum_{ \substack{i\in \mathcal{T} \\ j \centernot \in \mathcal{T} }} J^{(to)}_{ij} s_is_j}_{H_{\mathcal{TO}}}, \label{spin_hamiltonian} \end{equation} 
 where  $H_{\mathcal{T}}$ and $H_{\mathcal{O}}$ are the Hamiltonian of the subsystems $S_{\mathcal{T}}$ and $S_{\mathcal{O}}$, respectively, while $H_{\mathcal{TO}}$ describes the interactions between these subsystems.

%While  the couplings are   originally  updated at discrete time steps,  
 Now we introduce the effective potential to obtain the distribution of $\bold{J}$ \footnote{Note the difference from  the model implementation of  \cite{Sakata2009} where couplings are   updated at discrete time steps.}. For it, we consider a continuous Langevin-type dynamics for the couplings 
 \begin{equation}
      \frac{d J_{ij}}{d\tau } =  -\frac{1}{N} \frac{\partial\, V
      }{\partial  J_{ij}}  + \frac{1}{\sqrt{N}}\,\xi_{ij} (\tau )
     \,, 
     \label{diffusion}
\end{equation} 
where $V= V \big(\bold{J}\big)$  is the potential  of all the couplings and  $\xi_{ij}$ is the white noise whose intensity equal to the temperature  $T_J$. The factors $1/N$ and $1/\sqrt{N}$ in front of the potential and the noise term, respectively, ensure a correct relationship between the drift and diffusive parts  of the  Langevin equation. If the couplings were independent from each other, the potential would simply take the form of the potential of a free Brownian particle 
 \begin{equation}
   V_0 =  \frac{N}{2}\cdot \sum_{i<j} J_{ij}^2 \,.
     \label{free_particle}
\end{equation}
However, in the presence of fitness we need an additional term. 
Here we assume that the fitness would be maximised if a global alignment is established among target spins, so that we introduce   
\begin{equation}
\Psi = \frac{1}{N_t}\, \Big|\sum_{i \in \mathcal{T}} s_i  \Big|
\,, \label{fitness}
  \end{equation}
where $N_t$ is the size of $\mathcal{T}$.
Under this fitness that favors the alignment of target spins, the couplings are necessarily subjected to a fitness field $K$ 
\footnote{Technically speaking, the field $K$ is introduced to break 
the gauge-symmetry between different local Mattis states that have no frustration among target spins (i.e. triples $J_{ij}^{(tt)} J_{jk}^{(tt)} J_{ki}^{(tt)} > 0$, $\forall\, i<j<k \in \mathcal{T}$). As this symmetry is broken, the ferromagnetic state (with all links between target spins being positive) is distinguished from  all other unfrustrated states and, as a consequence, alignment among target spins can be achieved at low $T_s$. One might want to consider a different form of $K$ that explicitly takes into account the Boltzmann weights corresponding to the Hamiltonian $H_S$ (i.e. $e^{-\beta_s H_S}$, as defined in Eq. \eqref{fitness}).  Such a term, however, may not guarantee the evolution towards a subgraph of only ferromagnetic interactions  among target spins.
}: 
 \begin{equation}K =  \frac{1}{\beta_J}\frac{\partial}{\partial J_{ij} }\ln\Big( \sum_{\{s_i\}} \exp\Big\{ \frac{\beta_J}{N_t}\sum_{i<j \in \mathcal{T}} J_{ij}  \Big|\sum_{i \in \mathcal{T}} s_i  \Big|\Big\}\Big)
 \label{fitness_field}
  \end{equation}
  or equivalently, $V$ needs to be modified from $V_0$ into
   \begin{eqnarray*} \nonumber
  V =  V_0- \frac{1}{\beta_J}\, \ln\Big( \sum_{\{s_i\}} \exp\Big\{ \frac{\beta_J}{N_t}\,\sum_{i<j \in \mathcal{T}} J_{ij} \Big|\sum_{i \in \mathcal{T}} s_i  \Big|\Big\}\Big)\,.
     \label{free_particle_fitness}  
   \end{eqnarray*}
  Without the evolution of genotypes,   the Hamiltonians $H_{\mathcal{T}}$,  $H_{\mathcal{O}}$ and $H_{\mathcal{TO}}$ dictate the spins to adapt to a set of fixed couplings $\bold{J}$ in order to minimise each term of Eq. \eqref{spin_hamiltonian} through the spin dynamics.  Such adaptation  results in  an accordance between the state of $J_{ij}^{(tt)}$ and $s_is_j$ for $i,j \in \mathcal{T}$; that between the state of $J_{ij}^{(oo)}$ and  $ s_is_j $ for $i,j \centernot \in \mathcal{T}$ and that between the state of $J_{ij}^{(to)}$ and  $ s_is_j $ for $i \in \mathcal{T}, j \centernot \in \mathcal{T}$. As long as this kind of accordance exits, it is insufficient to consider the evolving couplings with selection force given in Eq. \eqref{fitness_field}  only. A link between two spins hence necessarily needs to adapt to the joint state of these spins. Due to the time scale separation between the phenotype- and the genotype dynamics, %the dependence of $J_{ij}$ on $s_i s_j$ should take the form
  the direction of change of genotypes is determined by the equilibrium correlation of the phenotypes, and since $J_{ij}$ is symmetric, it needs to be:
  \begin{equation*} 
    d J_{ij}/d\tau  \propto \langle s_is_j \rangle_{T_s} \,.
  \end{equation*}
This is equivalent to have a potential of the form \footnote{Here we used the identities $\langle s_is_j \rangle_{T_s}  = \frac{1}{
  \beta_s } \frac{\partial }{\partial  J_{ij}^{(tt)}}  \ln Z_1$ and $\langle s_ks_{k'} \rangle_{T_s}   = \frac{1}{ \beta_s } \frac{\partial }{\partial  J_{kk'}^{(oo)}}  \ln \tilde{Z}_1$, where   the partition function of the target spins' subsystem and that of the non-target spins' subsystem are $Z_1:= \sum_{\{s_i\}_{i \in \mathcal{T}}} e^{-\beta_s  H_{\mathcal{T}} (\bold{J}^{(tt)})}$  and $\tilde{Z}_1:= \sum_{\{s_i\}_{i\centernot \in \mathcal{T}}} e^{-\beta_s   H_{\mathcal{O}} (\bold{J}^{(oo)})}$, respectively.}
   \begin{equation} 
  V_a =  V- \frac{1}{\beta_s}\, \ln\Big( \sum_{\{s_i\}} \exp\Big\{\beta_s\,\sum_{i<j} J_{ij} s_i s_j \Big\}\Big)  
     \label{free_particle_fitness_spin}  
   \end{equation}
The stochastic process induced by Eq. \eqref{diffusion} under this potential admits an equilibrium-like stationary   \emph{joint}  distribution  $\mathbb{P}(\bold{J}^{(tt)}, \bold{J}^{(oo)}, \bold{J}^{(to)})$ of Boltzmann-form (with associated temperature $T_J$), i.e., $$\mathbb{P}(\bold{J}^{(tt)}, \bold{J}^{(oo)}, \bold{J}^{(to)}) = e^{- \beta_J V_a}/\mathcal{Z}_{\rm total}\,, $$ 
where $\mathcal{Z}_{\rm total} = \sum_{\{\bold{J}\}} e^{-\beta_J V_a}$.  Instead of calculating this distribution, we introduce our approximate approach, in which  $J_{ij}^{(to)}$  are assumed to always attain equilibrium well before  $J_{ij}^{(tt)}$ and $J_{ij}^{(oo)}$ and hence can be adiabatically eliminated. As a consequence, only  the  weights of  equilibrium configurations of $\bold{J}^{(to)}$ contribute to the stationary  distributions
\begin{eqnarray*}
\mathbb{P}_{\mathcal{T}}(\bold{J}^{(tt)}) &=& \lim_{\tau \rightarrow \infty}  \mathbb{P}_{\mathcal{T}}(\bold{J}^{(tt)},\tau) \\ \mathbb{P}_{\mathcal{O}}(\bold{J}^{(oo)}) &=& \lim_{\tau \rightarrow \infty}  \mathbb{P}_{\mathcal{O}}(\bold{J}^{(oo)},\tau)
\end{eqnarray*}
where $\mathbb{P}_{\mathcal{T}}(\bold{J}^{(tt)},\tau)$ and $\mathbb{P}_{\mathcal{O}}(\bold{J}^{(oo)})$ are the 
 time-dependent solutions of the corresponding Forrker-Planck equations % that are associated to Eq. \eqref{diffusion_new} 
 with the effective potentials $V_{tt}$ for  $\bold{J}^{(tt)}$  and $V_{oo}$ for and $\bold{J}^{(oo)}$, respectively \footnote{Requiring that the couplings should remain bounded as $\tau \rightarrow \infty$, we need to introduce a decay term $-\lambda_{ij} J_{ij}$ to the couplings dynamics. In order to keep the genotypic selection pressure the same for all the couplings regardless of their types, we choose $-\lambda_{ij} J_{ij}^{(tt)} = -\sqrt{p}  J_{ij}^{(tt)}$ and $-\lambda_{ij} J_{ij}^{(oo)} = -\sqrt{1-p}  J_{ij}^{(oo)}$, where $p = N_t/N$ is the fraction of target spins}. %; $N_{t, t} = N_t$ and $N_{t, o} = N-N_t$.   
 To  obtain the distribution only of $\bold{J}^{(tt)}$ and $\bold{J}^{(oo)}$, we first replace $\bold{J}^{(to)}$ as given matrix by that obtained self-consistenly from equilibrium distribution. For it we need to modify $V_a$ in such a way that  the influence of $\bold{J}^{(to)}$ on $\bold{J}^{(tt)}$ ($\bold{J}^{(oo)}$) can be taken into account in the effective potential  $V_{tt}$ ($V_{oo}$). The \emph{joint}  effect  of $H_{\mathcal{TO}}$ and $H_{\mathcal{T}}$  on  the dynamics of target spins  suggests  that the dependence of  $J_{ij}^{(tt)}$ and $J_{ik}^{(to)} $ on each other arises from  any triad formed between $(i,j)\in \mathcal{T}$ and $k \centernot \in \mathcal{T}$, i.e., via $J_{ij}^{(tt)} J_{ik}^{(to)} J_{jk}^{(to)}$. This influence is represented by the  frustration \cite{Mezard} which implies that an optimal spin configuration $(s_i^*, s_j^*, s_k^*)$ can only be established  if the relation $J_{ij}^{(tt)} J_{ik}^{(to)} J_{jk}^{(to)} >0$ holds (optimality in this context means that $H_{\mathcal{TO}}$  and $H_{\mathcal{T}}$ can be lowered simultaneously by $(s_i^*, s_j^*, s_k^*)$).   Since for any given pair of $(i,j) \in \mathcal{T}$ there are  $N- N_t$ triads $\Delta_k$ formed between it and a non-target spin $k \centernot \in \mathcal{T}$, the total effect of frustration is given by
\begin{equation}
   F_{ij}^{(t-o-t)} = (N-N_t)^{-1}\sum_{k =1}^{N-N_t}  J_{ik}^{(to)}J_{jk}^{(to)} \,.
   \label{tot_triad_term}
   \end{equation}
Likewise, frustration among all $N_t$ triads $\tilde{\Delta}_k$ formed between a given pair of  non-target spins $i\centernot \in \mathcal{T}$ and $j\centernot \in \mathcal{T}$  with  $k \in \mathcal{T}$ induces a force   $ F_{(o-t-o)}$ on the state of $J_{ij}^{(oo)}$:
   \begin{equation} 
 F_{ij}^{(o-t-o)} = N_t^{-1}\sum_{k =1}^{N_t}  J_{ik}^{(to)}J_{jk}^{(to)}\,.
  \label{oto_triad_term}
   \end{equation}
The proposed scheme just allows us to define % an equilibrium distribution for  all the $\bold{J}^{(to)}$ based on a Hamiltonian\begin{equation}H^{(to)} =  -  \sum_{i<j\in \mathcal{T}} J_{ij}^{(tt)}  F_{ij}^{(t-o-t)}  - \sum_{k<k'\centernot \in \mathcal{T}} J_{kk'}^{(oo)}  F_{kk'}^{(o-t-o)}\label{Hamiltonian_for_Jto} \end{equation} with  $\bold{J}^{(to)}$ being the degrees of freedom. Now we can compute the equilibrium statistics of $\bold{J}^{(to)}$ with the help of the Boltzmann distribution $\mathbb{P}^{(to)}$ associated to  $H^{(to)}$.Let $\overline{J_{ik}^{(to)} J_{jk}^{(to)}}$ denote  the correlations   between those pairs of $J_{ik}^{(to)} $ and $J_{jk}^{(to)}$ that  have a common non-target spins $k$ according to $\mathbb{P}^{(to)}$. T
the effective potential $V_{tt}$ for  $\bold{J}^{(tt)}$  and $V_{oo}$ for and $\bold{J}^{(oo)}$ as
\begin{eqnarray*}
V_{tt} &=& V_a - \frac{1}{\beta_J}\, \ln\Big( \sum_{\{J_{ij}^{(to)}\}} \exp\Big\{\beta_J\,\sum_{i<j \in \mathcal{T}} J_{ij}^{(tt)} F_{ij}^{(t-o-t)} \Big\}\Big)  
\\ 
  V_{oo} &=& V_a - \frac{1}{\beta_J}\, \ln\Big( \sum_{\{J_{ij}^{(to)}\}} \exp\Big\{\beta_J\,\sum_{i<j \centernot \in \mathcal{T}} J_{ij}^{(oo)} F_{ij}^{(o-t-o)} \Big\}\Big) 
\end{eqnarray*}
 The stationary distributions induced by the diffusion process in Eq. \eqref{diffusion} with these effective potentials have a Boltzmann-form $
\mathbb{P}_{\mathcal{T}}(\bold{J}^{(tt)}) = e^{-\beta_J V_{tt}}/ \mathcal{Z}_{\mathcal{T}} $
and $\mathbb{P}_{\mathcal{O}}(\bold{J}^{(oo)}) = e^{-\beta_J V_{oo}}/ \mathcal{Z}_{\mathcal{O}} $
where  
 $\mathcal{Z}_{\mathcal{T}}$ and $\mathcal{Z}_{\mathcal{O}}$ are the partition function of  the genotypes $\bold{J}^{(tt)}$  and that of  the genotypes $\bold{J}^{(oo)}$, respectively \footnote{Note that $\mathcal{Z}_{\mathcal{T}}$ and $\mathcal{Z}_{\mathcal{O}}$  are different from the partition functions of the target spins' subsystem  $Z_1(\bold{J}^{(tt)})$ and that of the non-target spins' subsystem $\tilde{Z}_1(\bold{J}^{(oo)})$  given in [26].}. Here we replace $\bold{J}^{(to)}$  by the replica matrix $ \sigma_{i}^k :=J_{ik}^{(to)}$, to be obtained. (This stepwise scheme is valid, as we are concerned with the equilibrium property).  Denoting $n := T_s/T_J$, we can compute $\mathcal{Z}_{\mathcal{T}}$ and $\mathcal{Z}_{\mathcal{O}}$ as  %\footnote{To have a sensible thermodynamics limit for each of the subsystems $S_{\mathcal{T}}$ and $S_{\mathcal{O}}$, in these equations the couplings are required to be rescaled as $J_{ij}^{(tt)} \propto 1/\sqrt{N_t}$ and  $J_{ij}^{(oo)} \propto 1/\sqrt{N-N_t}$ . As a consequence of this rescaling, target and non-target spins are effectively coupled to heat baths at  different  temperatures $\beta_s^{(t)} = \beta_s/\sqrt{p}$ and $\beta_s^{(o)} = \beta_s/ \sqrt{1-p}$, respectively. %Without any further constrain, there is no restriction on the specific values of the scaling parameters $\lambda$ and $\sigma$ that should be used. If we want This is to ensure that the product of the coupling and inverse temperature  are the same for both $S_{\mathcal{T}}$ and  $S_{\mathcal{O}}$ as for the original system $S$, i.e. $ \beta_s^{(t)} J_{ij}^{(tt)} = \beta_s^{(o)} J_{ij}^{(oo)} = \beta_s J_{ij} $}
 
\begin{widetext}
 \[
 \mathcal{Z}_{\mathcal{T}}   = \int\prod_{i<j \in \mathcal{T}} dJ^{(tt)}_{ij} \sum_{\big\{s_i;s_i^\alpha;\sigma_i^k\big\}_{i \in \mathcal{T}}} \exp\left\{\beta_J \sum_{i<j}  \Big[- \frac{ N_t}{2} \, \Big(J_{ij}^{(tt)}\Big)^2 + J_{ij}^{(tt)}\Big(\frac{1}{N_t}\, \Big|\sum_{i \in \mathcal{T}} s_i\Big|  + \underbrace{\frac{1}{n}\,\sum_{a =1}^n s_i^a s_j^a}_{s-{\rm  replicas}} + \underbrace{\frac{1}{N-N_t}\, \sum_{k =1}^{N-N_t} \sigma_i^k \sigma_j^{k}}_{\sigma-{\rm replicas}}   \Big)\Big]  \right\} 
 \]
 \end{widetext}
 \begin{widetext}
  \[ \mathcal{Z}_{\mathcal{O}}   =  \int\prod_{i<j\centernot \in  \mathcal{T}} dJ^{(oo)}_{ij} \sum_{\big\{s_i^\alpha;\sigma_i^k\big\}_{i\centernot \in \mathcal{T}}} \exp\left\{\beta_J \sum_{i<j}  \Big[- \frac{N-N_t}{2} \, \Big(J_{ij}^{(oo)}\Big)^2 + J_{ij}^{(oo)}\Big(\tilde{K} + \underbrace{\frac{1}{n}\,\sum_{a =1}^n s_i^a s_j^a}_{s-{\rm  replicas}}   + \underbrace{\frac{1}{N_t}\, \sum_{k =1}^{N_t} \sigma_i^k \sigma_j^{k}}_{\sigma-{\rm  replicas}}    \Big)\Big]  \right\}  
  \]
   \end{widetext}
   In writing these equations, we assume that the fitness  acts only on $J_{ij}^{(tt)}$ \footnote{This can already be seen directly from Eq. \eqref{fitness_field} as the expression inside the bracket does not depend on $J_{ij}^{(oo)}$ and $J_{ij}^{(to)}$.} and hence in $\mathcal{Z}_{\mathcal{O}}$ we replace the fitness field $K$  by a constant $\tilde{K}$, which eventually will be set to 0  by virtue of calculations of the observables for non-target spins. Although neglecting the fitness's effect on $J_{ij}^{(oo)}$ does not follow exactly the above-mentioned implementation of the model, we expect that this holds true in the long times limit because otherwise both target- and non-target configurations at equilibrium would determine the fitness. This restriction hence corresponds to a first-order approximation of the fitness's effect, while  a  term that affects the dynamics of $J_{ij}^{(oo)}$ and $J_{ij}^{(to)}$  is considered  to be of higher order.

 We here propose to interpret $\sigma_{i}^k$ as the $k$-th replica of another variable $\sigma_i \in \{-1,1\}$ that is also located at the site $i$ of the graph (generally $\sigma_i \neq s_i$). To distinguish these different types of replica from each other, we call $s_i^a$ spin-replica and $\sigma_{i}^k$ coupling-replicas.   Following this interpretation, apart from  $n$ that appears as the number of spin-replicas $s_i^a$, $a=\{1,\cdots,n\}$, in  $\mathcal{Z}_{\mathcal{T}}$ and $\mathcal{Z}_{\mathcal{O}}$ \footnote{Though $n$ appears as integer number here, we will analytically continue to real positive $n$ in subsequent steps of calculations.}, we thus have $N-N_t$ coupling-replicas, $\sigma_i^k$, for $i\in \mathcal{T}$ and $k=\{1,\cdots,N-N_t\}$ and $N_t$ coupling-replicas, $\sigma_i^k$, for $i \centernot \in \mathcal{T}$ and $k=\{1,\cdots,N_t\}$ respectively. As in general, none of these numbers are zero, our double-replica approach does not correspond to the conventional quenched limit in spin-glass models \cite{Mezard}. Once setting $K = 0$ and neglecting the terms corresponding to $F^{(t-o-t)}$ and $F^{(o-t-o)}$, we recover  Coolen et. al. model for neural systems with dynamic synapses \cite{Coolen1993}. In contrast to the use of a Hamiltonian for the couplings adopted in \cite{Sakata2012}, 
here, we have  introduced the effective potential  for couplings that, by using the time-scale separation between the dynamics of genotypes and that of phenotypes, allows for the integration of the spin dynamics specified by the Hamiltonian Eq.  \eqref{spin_hamiltonian} into the Langevin dynamics of the couplings through the second term in Eq. \eqref{free_particle_fitness_spin}.   
 
Here we characterise the equilibrium behaviour of the model by the average fitness, $m$, the overlap between different spin replicas $a$ and $b$, $q_{ab}$, and the correlation between adjacent links  $Q$. Additionally, we want to quantify the mean value of the couplings  among the target spins only $\Phi$ . Let $\mathbb{E}[\cdot]$ and $\mathbb{\tilde{E}}[\cdot]$ denote ensemble average over  $\mathbb{P}_{\mathcal{T}}(\bold{J}^{(tt)})$ and $\mathbb{P}_{\mathcal{O}}(\bold{J}^{(oo)})$, respectively. These order parameters are given by 
 \begin{subequations}
\label{allequations2}
 \begin{eqnarray}
m_a = \mathbb{E}\Big[s_i^a\Big]_{i\in \mathcal{T}} \,,\,\,\, q_{ab} = \mathbb{E}\Big[s_i^a s_i^b\Big]_{i\in \mathcal{T}} 
  \label{orderparameter1}
\\
    Q_{kk'} = \mathbb{E}\Big[J_{ik}^{(to)} J_{ik'}^{(to)}\Big]_{ \substack{i\in \mathcal{T} \\ k,k' \centernot \in \mathcal{T} }} \label{orderparameter2}
    \\
   \Phi = \mathbb{E}\Big[J_{ij}^{(tt)} \Big]_{ \substack{i,j \in \mathcal{T} }}
       \label{orderparameter3}
\end{eqnarray}
\end{subequations}
Similarly, for the non-target spins we have
 \begin{subequations}
\label{allequations3}
 \begin{eqnarray}
\tilde{m}_a = \mathbb{\tilde{E}}\Big[s_i^a\Big]_{i \centernot\in \mathcal{T}} \,,\,\,\, \tilde{q}_{ab} = \mathbb{\tilde{E}}\Big[s_i^a s_i^b\Big]_{i\centernot\in \mathcal{T}} 
  \label{orderparameter4}
\\
    \tilde{Q}_{kk'} = \mathbb{\tilde{E}}\Big[J_{ik}^{(to)} J_{ik'}^{(to)}\Big]_{ \substack{i\centernot \in \mathcal{T} \\ k,k'  \in \mathcal{T} }}  \label{orderparameter5}
        \\
   \tilde{\Phi} = \mathbb{\tilde{E}}\Big[J_{ik}^{(oo)} \Big]_{ \substack{i,j \centernot \in \mathcal{T} }}
       \label{orderparameter6}
\end{eqnarray}
\end{subequations}
In the thermodynamics limit, $N\rightarrow \infty$ and $N_t\rightarrow \infty$, while keeping $p=N_t/N$ fixed,  using a replica symmetric ansatz for the variables $m_a = m$, $q_{ab} = q$ and $Q_{kk'} = Q$, $\tilde{m}_a = \tilde{m}$; $\tilde{q}_{ab} = \tilde{q}$ and $\tilde{Q}_{kk'} = \tilde{Q}$, as well as, $M_{ak} = M$ and $\tilde{M}_{ak} = \tilde{M}$, where  $M_{ak} = \mathbb{E}\big[s^a_i J_{ik}^{(to)}\big]_{ \substack{i \in \mathcal{T} \\ k\centernot  \in \mathcal{T} }}$ and  $\tilde{M}_{ak} = \mathbb{\tilde{E}}\big[s^a_i J_{ik}^{(to)}\big]_{ \substack{i \centernot \in \mathcal{T} \\ k  \in \mathcal{T} }}$, we obtain the following free energy densities:
 \begin{subequations}
 \label{allequations4}
 %\begin{align}
 \begin{eqnarray}
   \nonumber f^{\rm RS}_{\mathcal{T}} & = & 
   \frac{1}{2}   \,\left\{\frac{q}{n} + \frac{Q}{N-N_t}   +\frac{(n-1)q^2}{2n} +  \frac{ Q^2}{2} +  M^2\right\} \\ 
   &-&\frac{1}{\beta_J} \ln\left[\sum_{z=0}^{N_t} \binom{N_t}{z}   \sum_{k=0}^{N-N_t} \binom{N-N_t}{k} I_{kz}\right] 
\label{freeenergy1}
        \\
 \nonumber
 f^{\rm RS}_{\mathcal{O}} & = &
 \frac{1}{2}   \,\left\{\frac{ \tilde{q}}{n} + \frac{\tilde{Q}}{N_t}  +\frac{(n-1) \tilde{q}^2}{2n} +  \frac{\tilde{ Q}^2}{2} +   \tilde{M}^2\right\} \\ & -&\frac{1}{\beta_J}\ln \left\{ \sum_{k=0}^{N_t} \binom{N_t}{k} \tilde{I}_{k} \right\}
%\end{align}
   \label{freeenergy2}
\end{eqnarray}
\label{freeenergy}
\end{subequations}
 From the extremum condition of these free energies we can compute all the model order parameters via a set of self-consistency equations. These equations as well as the functions $I_{kz}$ and $\tilde{I}_k$ are given in the SM \footnote{Here $I_{kz} = I_{kz}(m,q,Q,r,M)$ and $\tilde{I}_k = \tilde{I}_{k} (\tilde{m},\tilde{q},\tilde{Q}, \tilde{r}, \tilde{M})$ with $r, M , \tilde{r}, \tilde{M}$ are other variables that the free energy densities depend on. Since these observables for target and non-target spins have their behaviour correlated to that of $(m,q,Q)$ and $(\tilde{m}, \tilde{q}, \tilde{Q})$, respectively, they do not provide additional information about the structure of the model phase diagram.}.  The use of the replica symmetric is justified in most part of the $(T_s, T_J)$ parameter space from the stability analysis \cite{deAlmeida_1978}.  At  low $Ts$ and $T_J$ the replica symmetry is broken, which, we will not explore fully. Nevertheless we will  discuss later how robustness of phenotypes, postulated for biological systems that reproduce similar offspring, is lost in that scenario. %Note \cite{Saakyan, Sakata2011}
 The  replica-symmetric free energy densities allow us to derive 
           \begin{subequations}
\label{allequations8}
 \begin{eqnarray}
\Phi & = & \frac{1}{N_t} \left[\Psi +  m^2 + r^2 \right]
   \label{mean_of_target_links}
\\
 \tilde{\Phi} &=&   \frac{\tilde{m}^2 +\tilde{r}^2}{N - N_t}   \,.
    \label{mean_of_non-target_links}
\end{eqnarray}
\end{subequations}
where $r$ and $\tilde{r}$ are defined and computed in the SM.
 \section{Phase diagram}
\begin{figure}[t]
\includegraphics[scale=0.215]{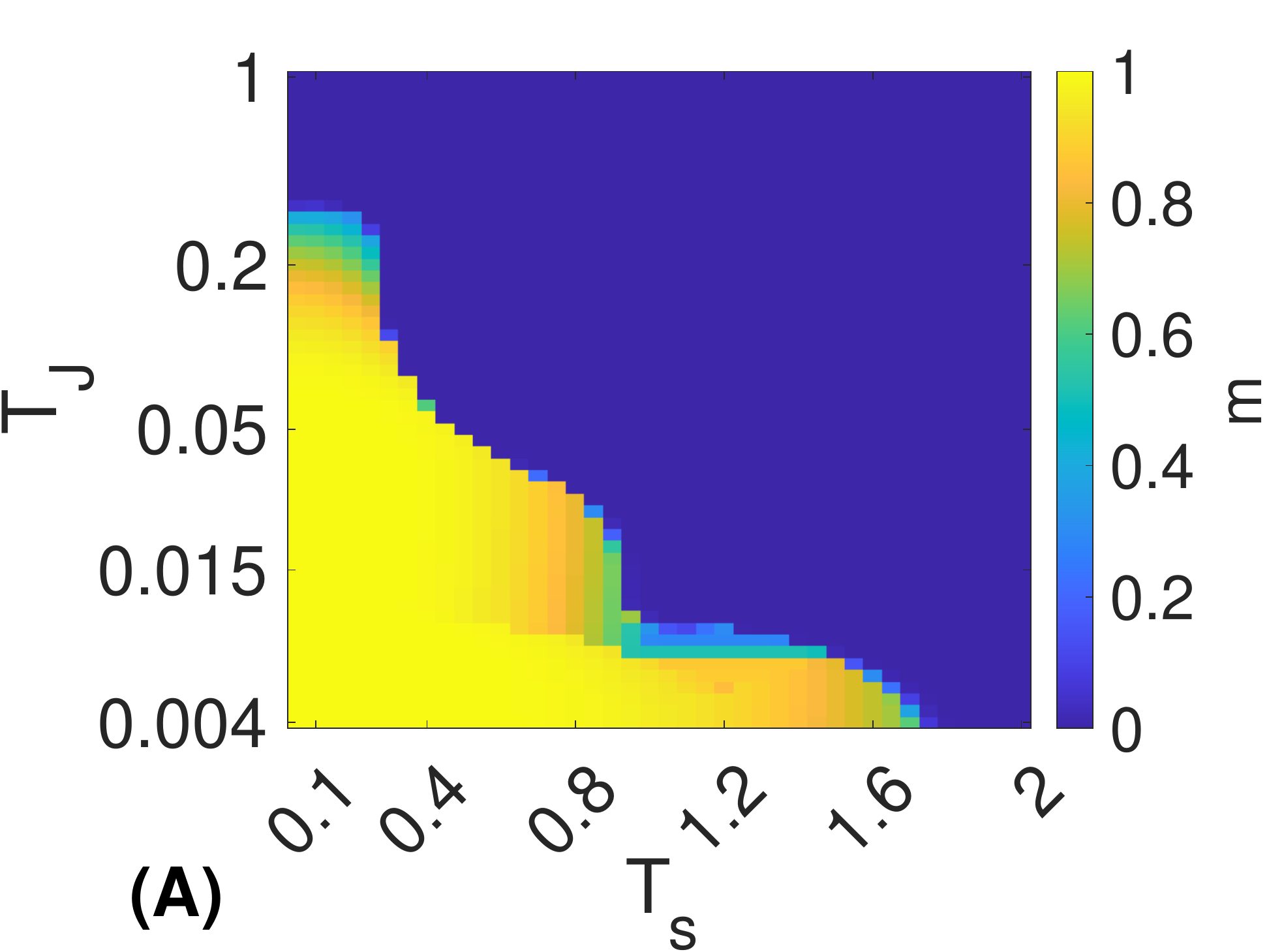}
\includegraphics[scale=0.215]{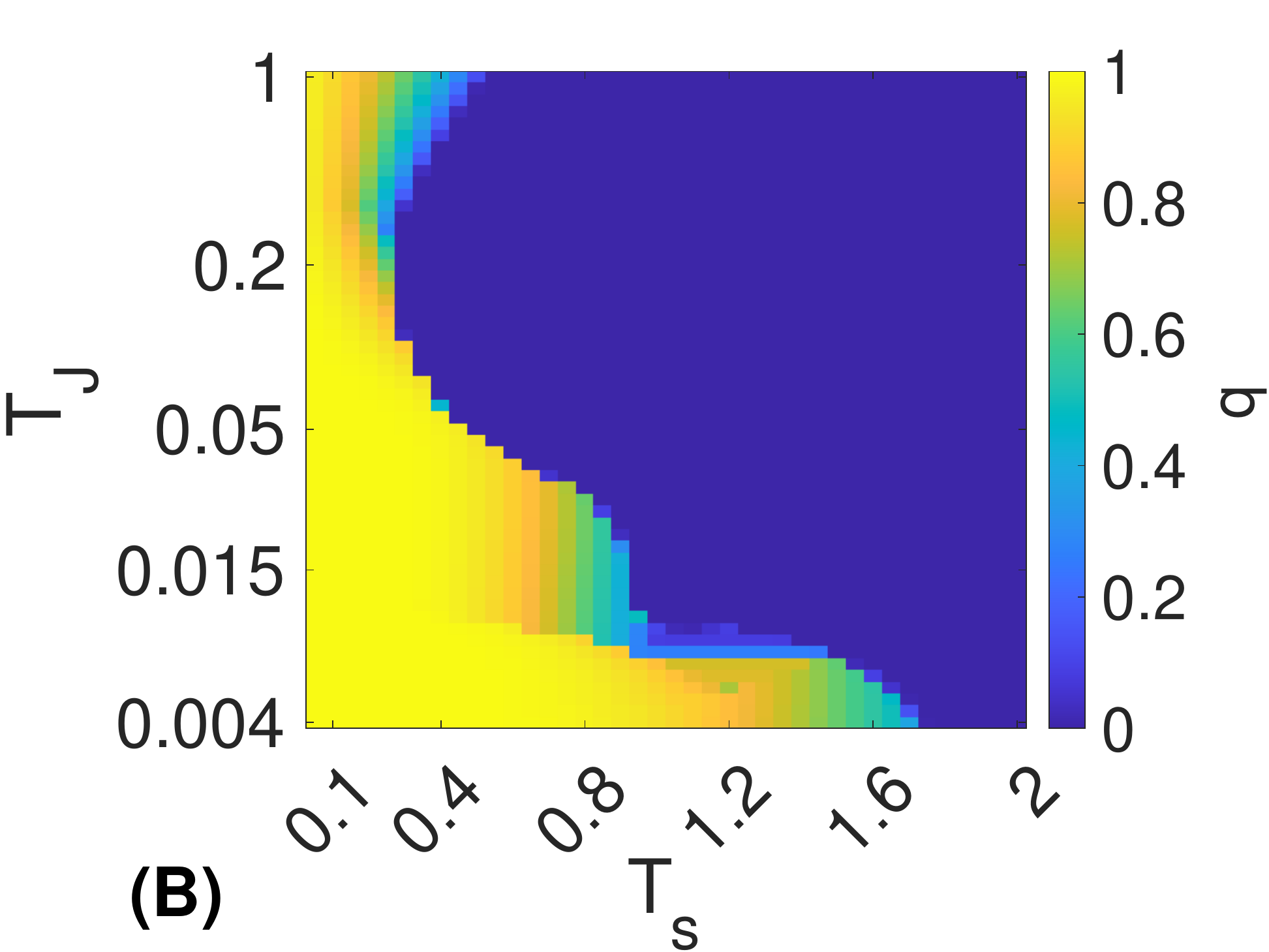}
\includegraphics[scale=0.215]{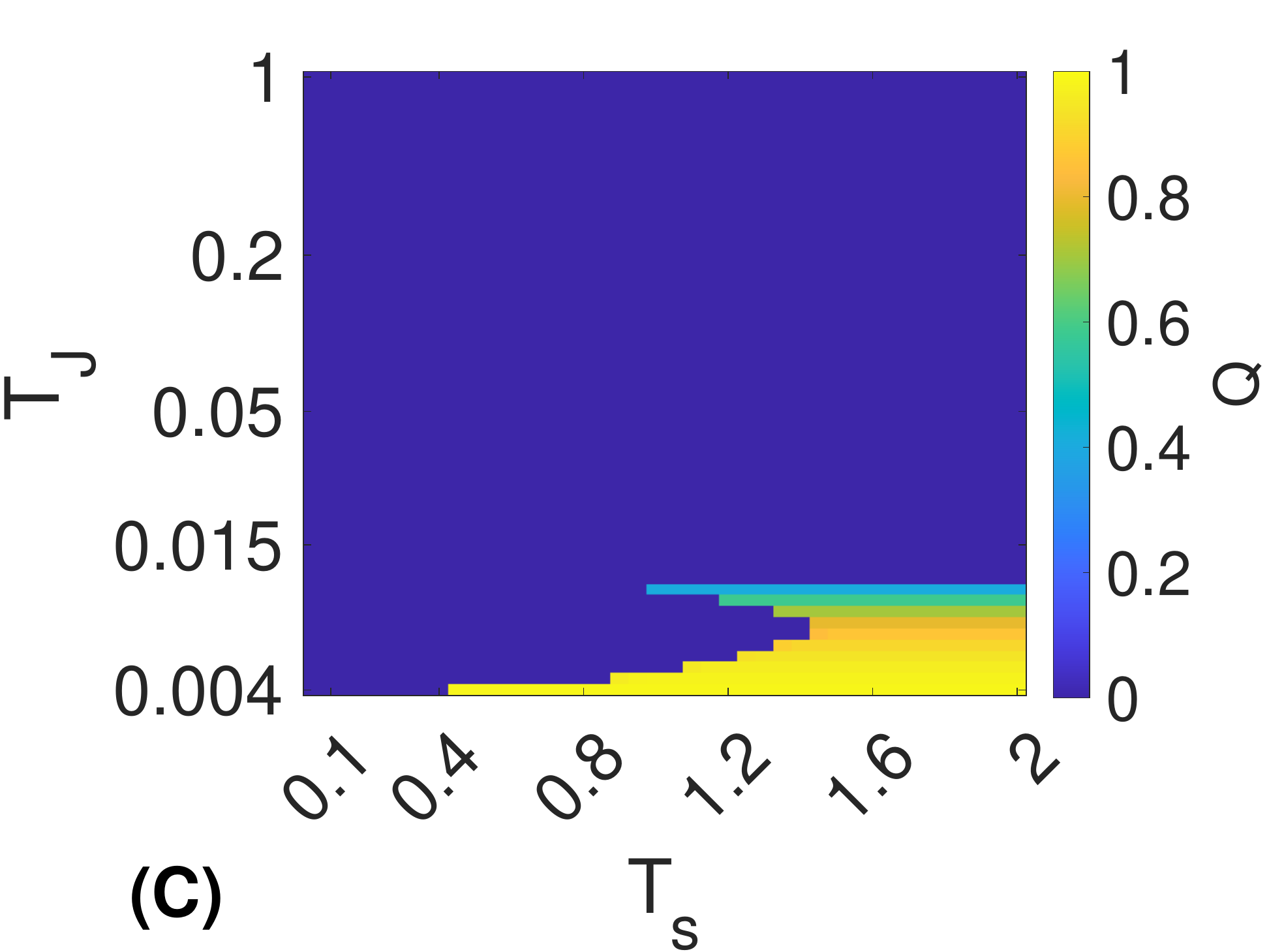}
\includegraphics[scale=0.215]{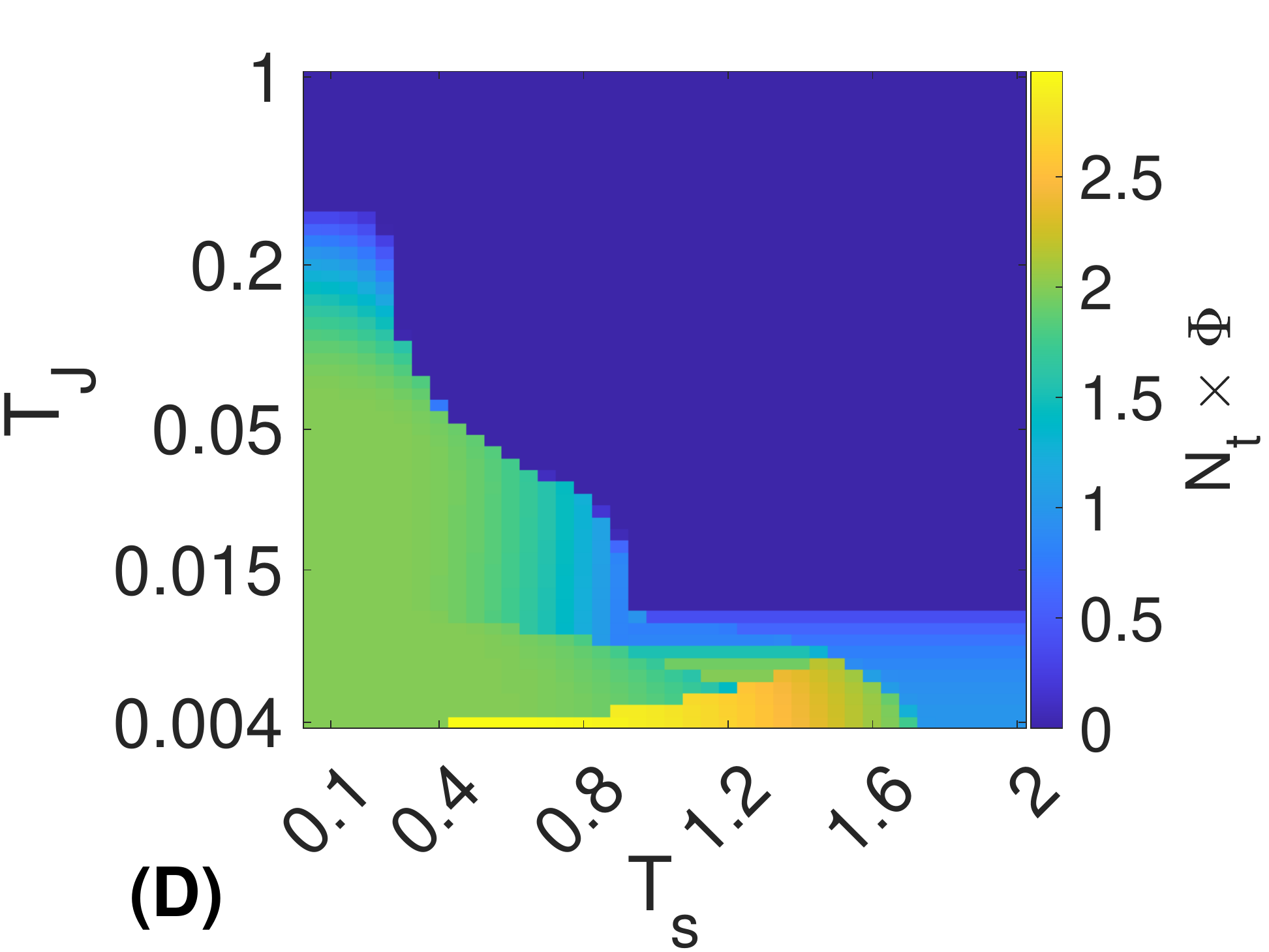}
\caption{Magnetisation  for target spins $m$ \textbf{(A)}. Overlap  between different replicas for target spins $q$ \textbf{(B)}. Averaged  correlation  of a pair of couplings between a target and a non-target spin that share a common non-target spin $Q$ \textbf{(C)}. Averaged frustration among target spins $\Phi$ \textbf{(D)}.  Here $N_t = 10$, $N = 100$. Note the $y$-axis is on logarithmic scale.} 
\label{fig:fig1}
\end{figure}
In Fig. \ref{fig:fig1} we depict the order parameters as function of the temperature $T_s$ and $T_J$ for a particular choice of $N=100$ and $N_t = 10$. Here for each point $(T_s, T_J)$ of the phase diagram, we solve numerically the set of mean-field equations for $m, q,Q$, while computing $\Phi$ from the knowledge of these quantities.  In   terms of only the magnetisation  $m$ and the overlap between spin-replicas $q$ for target spins, we observe  three distinct phases that are typical for spin-glass systems, namely, $m = q = 0$ (paramagnet phase);  $m = 0, q > 0$ (spin-glass phase),  and $m >0, q > 0$ with $\sqrt{q} > m$, ({ target-ferromagnet phase, 't-ferro' in short}). The transitions between the phases are second-order at small $T_s/T_J$, but  become discontinuous (first-order) at large $T_s/T_J$. At a much  lower value of $T_J$ there is a region where, apart from having a non-zero magnetisation of target spins, the order parameter $Q = \langle J^{(to)} J^{(to)} \rangle$ starts to become non-zero.  As can be anticipated from  Eq. \eqref{mean_of_target_links}, the mean value of $J_{ij}^{(tt)}$ also varies from region to region in accordance to  the change of  $m$ and that of $Q$. Note that on sharp contrary to the   transition between paramagnet and spin-glass which is similar to that of the Sherrington-Kirpatrick (SK) model, the phenotype-genotype coupling  results in a repositioning of the boundary between spin-glass
and t-ferro. Such difference arises from the non-zero correlation of the genotypes. Expanding the free energy $f^{\rm RS}_{\mathcal{T}}$ for small $m$ and $q$, the transition between  paramagnet and t-ferro occurs at $T_s^{\bold{P}\rightarrow\bold{F}} = \kappa$, where $\kappa = 2^{-N_t}\sum_{z=0}^{N_t} \binom{N_t}{z} |N_t-2z|/N_t$,\linebreak while   the spin-glass to t-ferro transition occurs at
$ \big[1+(n-1) q (\beta_s^{\bold{SP}\rightarrow\bold{F}} )  \big] \cdot \big(\beta_s^{\bold{SP}\rightarrow\bold{F}} \kappa  \big) =1$. We also check that both the magnetisation $\tilde{m}$ of  the non-target spins and the average value $\tilde{\Phi}$ of $J_{ij}^{(oo)}$  are always  zero as the non-target spin subsystem remains frustrated all the time, while the spin overlap $\tilde{q}$ can undergo a transition from paramagnet to spin-glass, in the same way as the SK model.  The phase diagrams of these quantities are given in the SM.

\begin{figure}[t]
\includegraphics[scale=0.3]{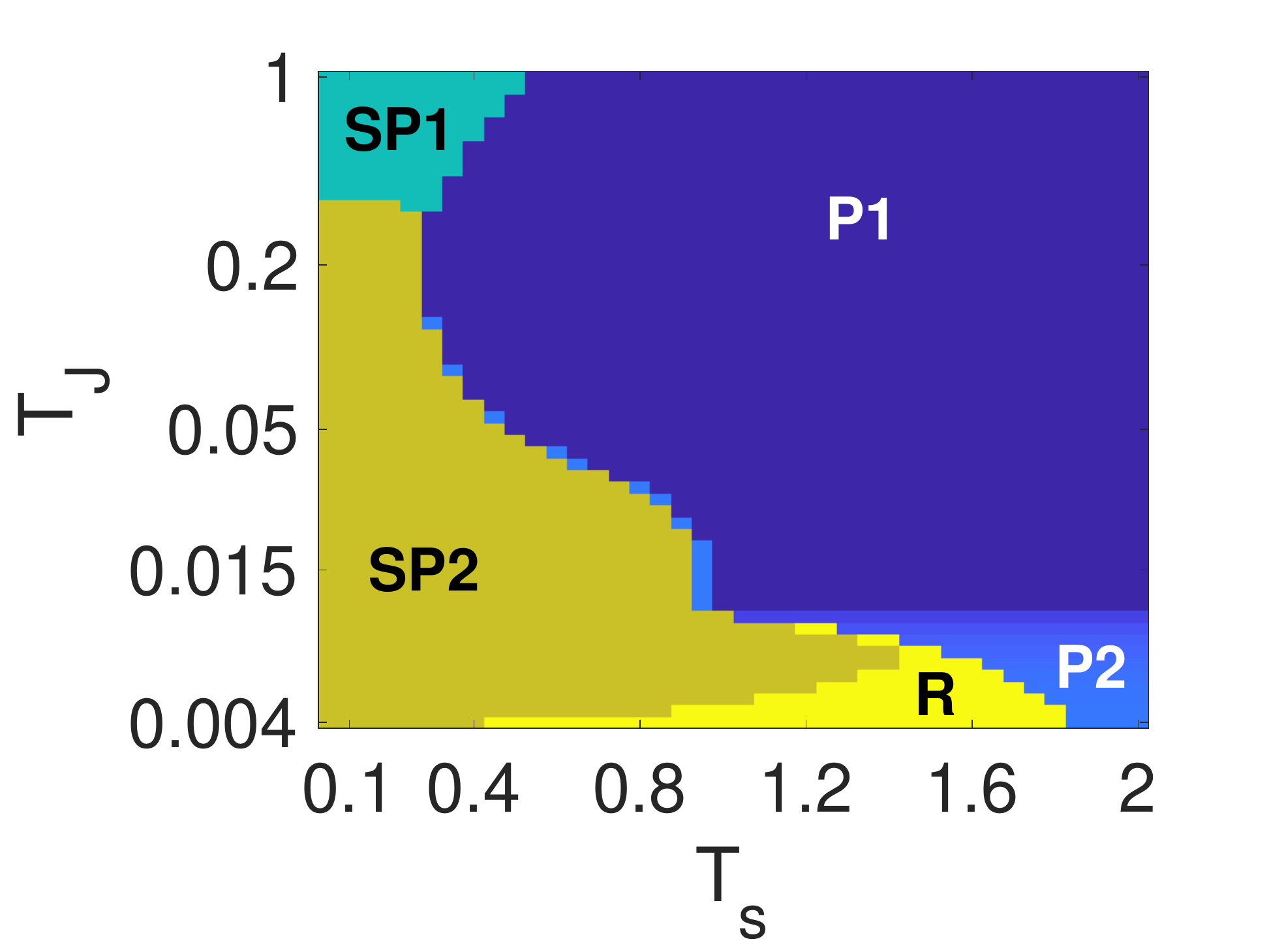}
\caption{The model phase diagram. Here \textbf{SP1} denotes the spin-glass phase with $m=\tilde{m} = Q= \tilde{Q} = 0$ and $q, \tilde{q} >0$;  \textbf{SP2} denotes the spin-glass phase with $\tilde{m} = Q= 0$ and $q, m >0$ (within this region, both $\tilde{q}$ and $\tilde{Q}$ can be either zero or non-zero, see SM); \textbf{P1} denotes the paramagnet phase with $m=\tilde{m} = q= \tilde{q} = Q = \tilde{Q} = 0$;  \textbf{P2} denotes the paramagnet phase with $m=\tilde{m} = q= \tilde{q} =  \tilde{Q} = 0$ but $Q > 0$; \textbf{R} denotes the robust fitted phase with $\tilde{m} =  \tilde{q} =  \tilde{Q} = 0$ but $m,q,Q >0$.  Here $N_t = 10$, $N = 100$. Note the $y$-axis is on logarithmic scale.} 
\label{fig:fig2}
\end{figure}
Combining the behaviour of the order parameters altogether, we obtain the model phase structure in Fig. \ref{fig:fig2}. It contains five distinct regions. At low genotypic selection pressure $T_J \geq e^{-1}$,  only the first spin-glass \textbf{SP1} and the paramagnet \textbf{P1} phases with zero fitness are observed. However, as the genotypic selection pressure  increases other phases emerge. At sufficiently low $T_J$, a robust fitted phase denoted by \textbf{R} ($m,q,Q > 0$) emerges in an intermediate range of $T_s$ (here $\tilde{m} = \tilde{q} = \tilde{Q} = 0$). Adjacent to this phase on the side of high phenotypic noise is the second paramagnet phase \textbf{P2} where the fitness value is low ($m = q = 0$) but there exists some structure in the genotypes such that  $Q > 0$.  On the other hand, for lower $T_s$,  the system is in the second spin-glass phase \textbf{SP2} with high
fitness  but non-robust genoptypes     ($m \simeq q \simeq 1, Q = 0$).  In particular, the transition from \textbf{R} to  \textbf{SP2}   is marked by a replica symmetry breaking (RSB) which indicates the loss of stability of the replica symmetric (RS) solutions \footnote{The eigenvalue of the Hessian changes its sign on the border between these phases, see SM. Note that though we do not calculate the full hierarchy of replica symmetry breaking (RSB) 
\`{a} la Parisi, but as long as RSB happens this non-robustness is true for the full RSB solution.}. The broad distribution of gene-gene correlations  in the RSB phase implies that the genotype of offsprings is not preserved, in contrast to the RS phase. In the biological context, this means that replication is no longer stable so that genotypes are not conserved over generations. 

In overall, the phase diagram agrees with what was observed numerically in \cite{Sakata2009, Sakata2009PRE}. However, thanks to the explicit account of the  coupling-replicas, so that $Q$ can be treated as an order parameter upon which the free energy density depends,    we discover the existence of the second paramagnet phase \textbf{P2} that was not reported before. This phase can be interpreted as a \emph{precusor} region, in which genotypes are structured in such a way that supports ferromagnetic ordering among target spins, and hence have  potentiality  to acquire  a high fitness, but due to the high fluctuation induced by $T_s$, this fitness cannot be maintained. Furthermore,  by considering separately the effective dynamics of the target and non-target subsystems, $S_{\mathcal{T}}$ and $S_{\mathcal{O}}$, our approach can differentiate the phase \textbf{SP1} from \textbf{SP2}. The previous approach \cite{Sakata2012}  only  stressed the distinct arrangement of target spins in the \textbf{SP2} region, where the subsystem of target spins becomes ferromagnetic whereas that of non-target ones remains spin-glass. Our present approach shows that this is no longer true for a high value of $T_J$. Upon increasing $T_J$, this ferromagnetic ordering is destroyed by genotypic fluctuations.

 \begin{figure}[t]
 \centering
   \includegraphics[scale=0.245]{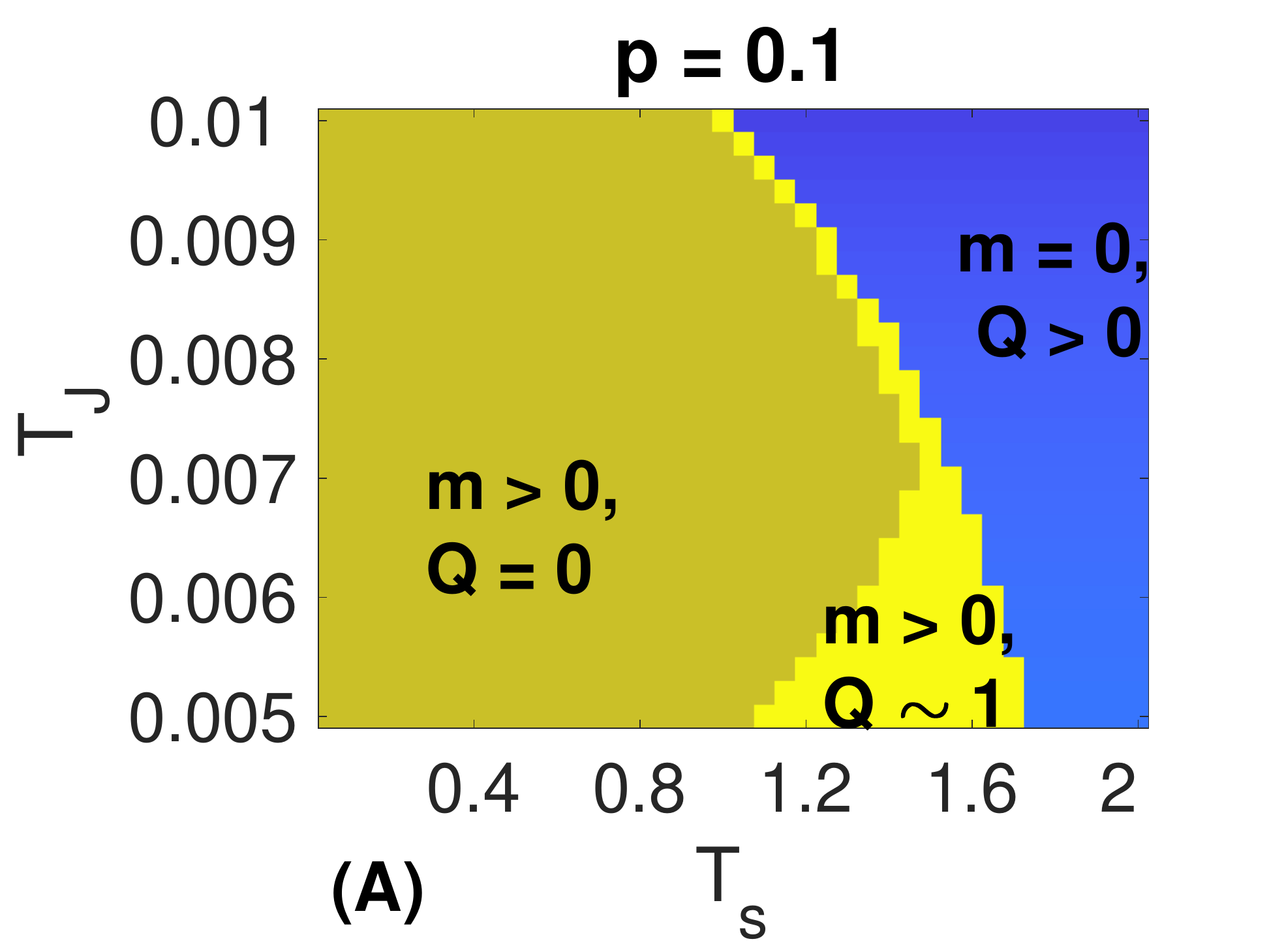}
      \includegraphics[scale=0.245]{ 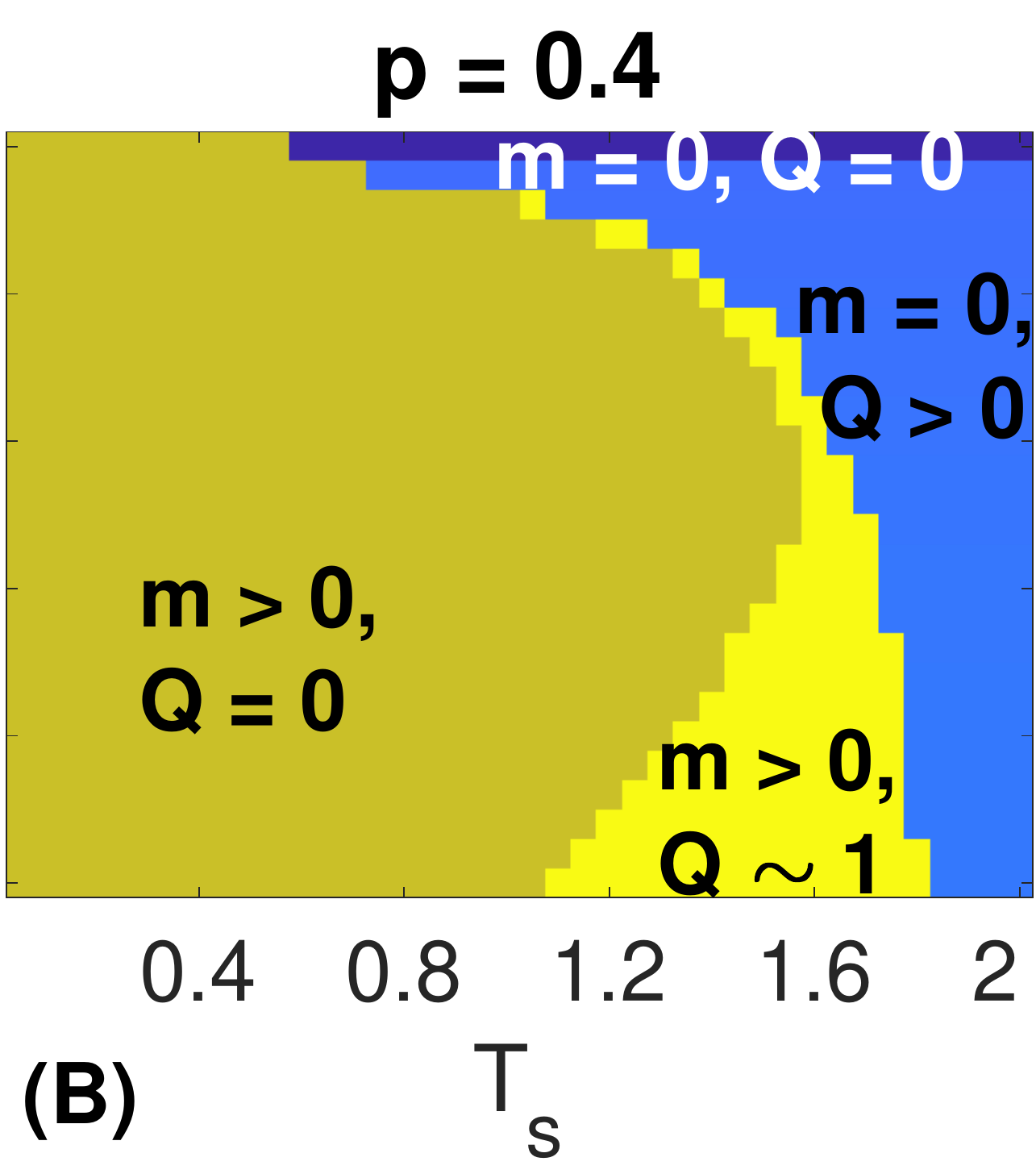}

         \includegraphics[scale=0.245]{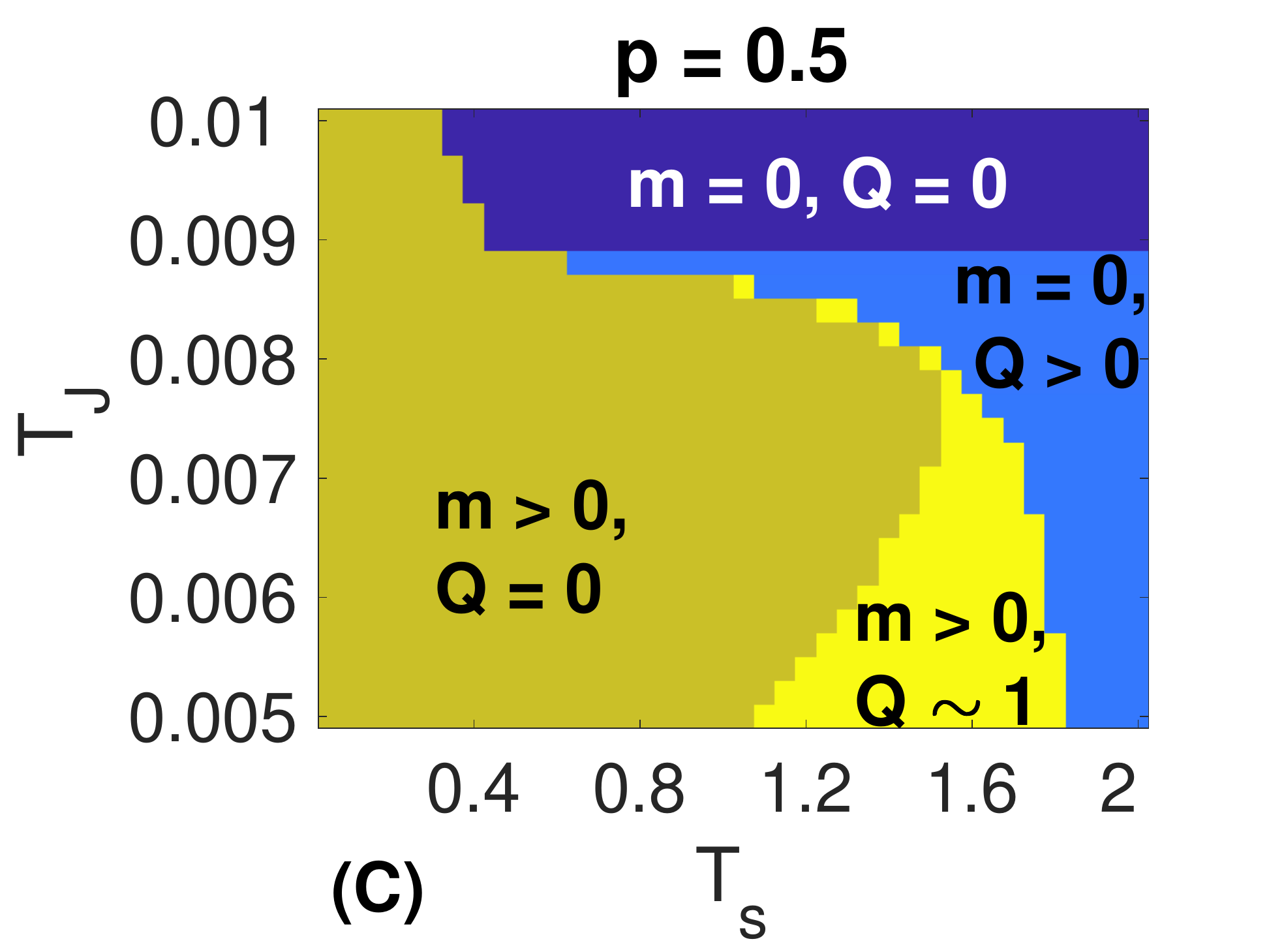}
          \includegraphics[scale=0.245]{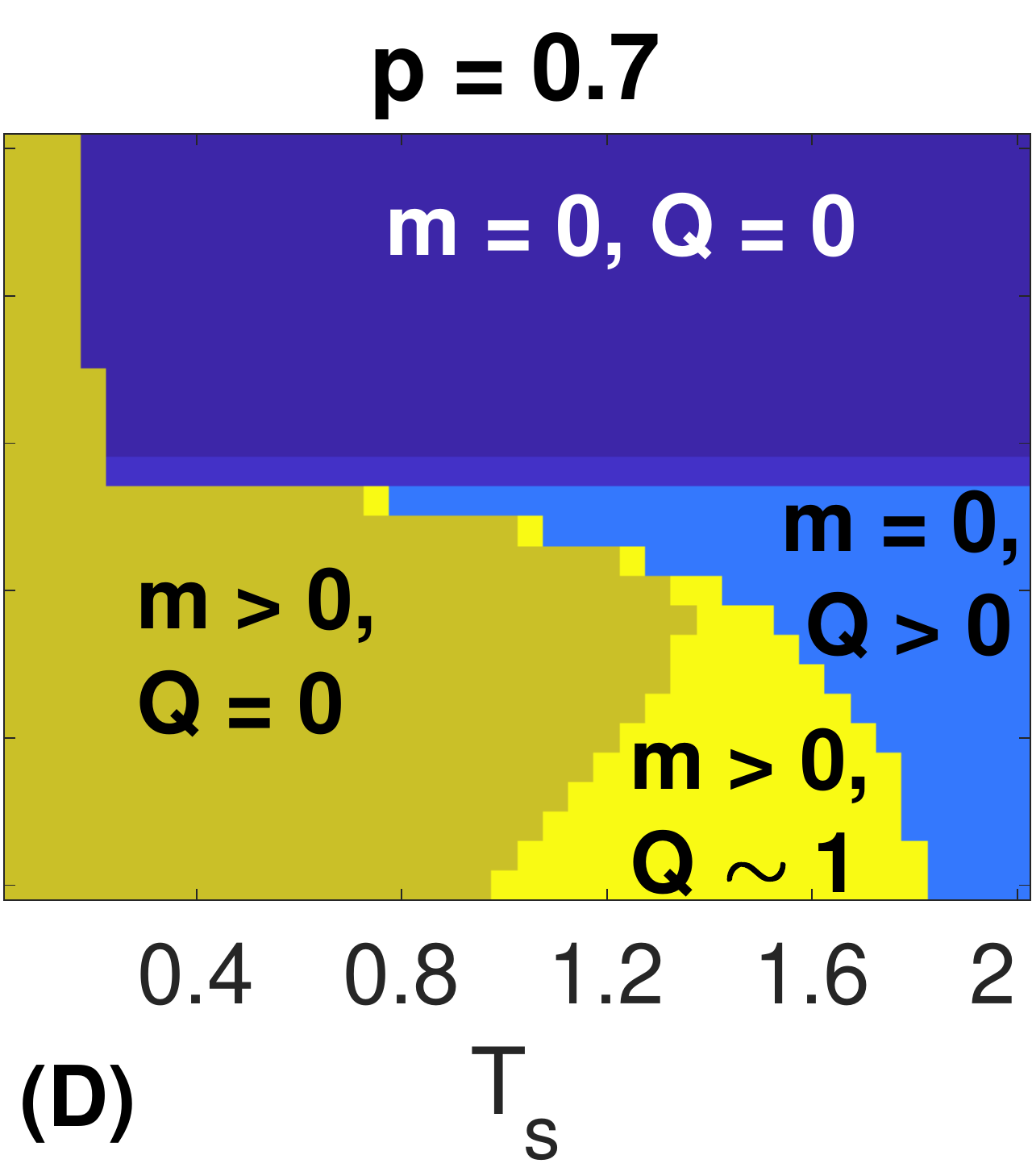}
\caption{The phase diagrams
obtained by combining the behaviour of $m$ as function of $T_s$ and $T_J$ with that of $Q$ for different values of $p$, at sufficiently low $T_J$. Here $N = 100$.} 
\label{fig:fig3}
\end{figure}
In addition, the present analysis allows one to obtain quantitative dependence of genotypic and phenotypic robustness on the fraction of targets.  The phase diagram in Fig. \ref{fig:fig2} includes global information of the system including weak selection region without achieving nonzero fitness, $m$ of target, whereas, of biological interest is if the fitted state is evolved robustly by the selection. To this end we focus on the low $T_J$ region of the phase diagram to explore the dependence of the system behaviour on the fraction $p= N_t/N$ of target spins. While overall, the phase  structure   is similar for different $p$ in Fig. \ref{fig:fig3}, in particular, the robust fitted (yellow) region seems to change slightly with increasing $p$, the relative size and exact location of all the other phases vary with $p$.   This suggests that a more quantitative analysis is needed  to understand the genotype-phenotype relationship as function of $p$. We carry on this analysis in the next section.   

 \section{ Structure of the robust fitted phase}
\begin{figure}[t]
    \includegraphics[scale=0.215]{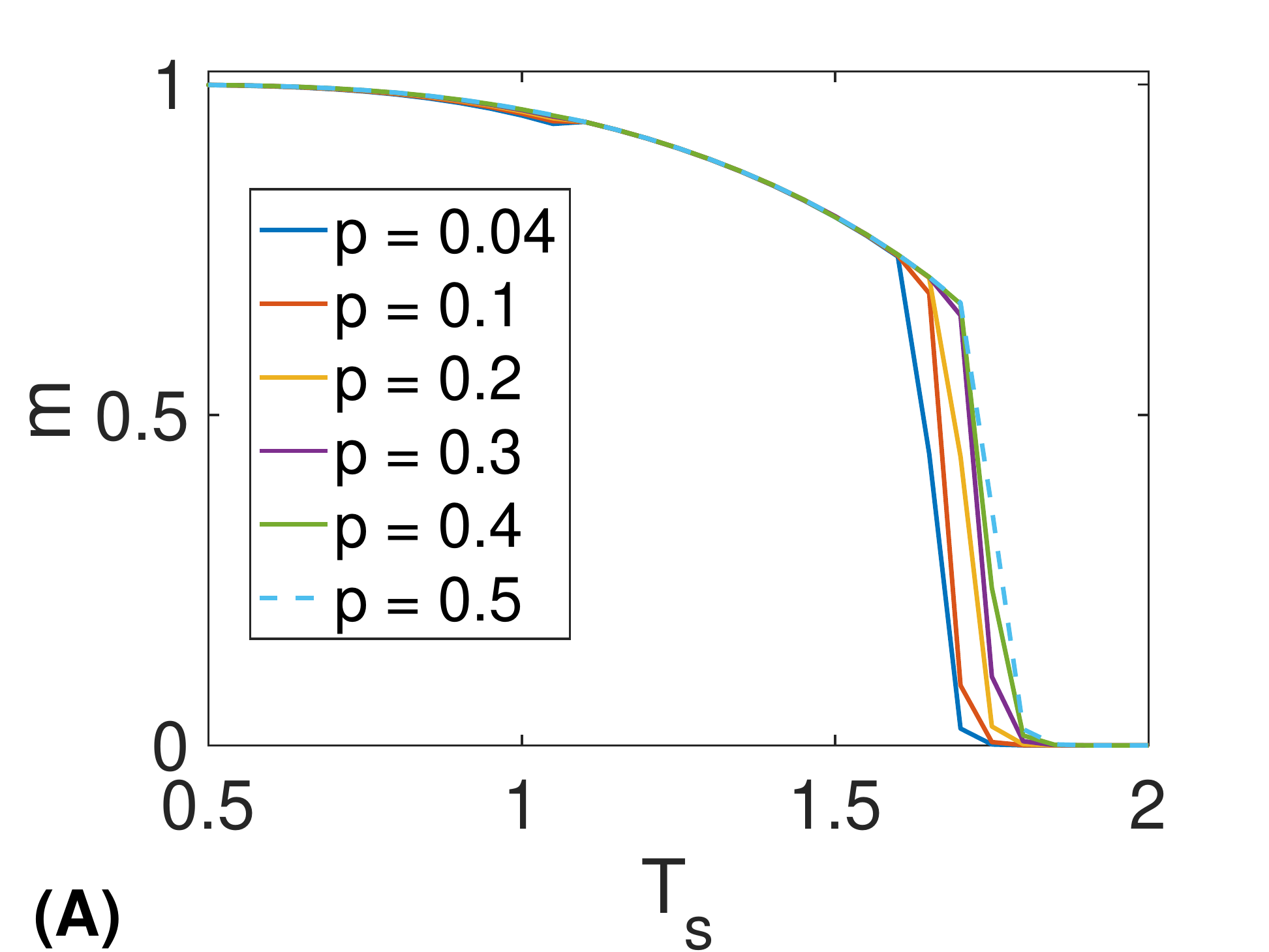}  
        \includegraphics[scale=0.215]{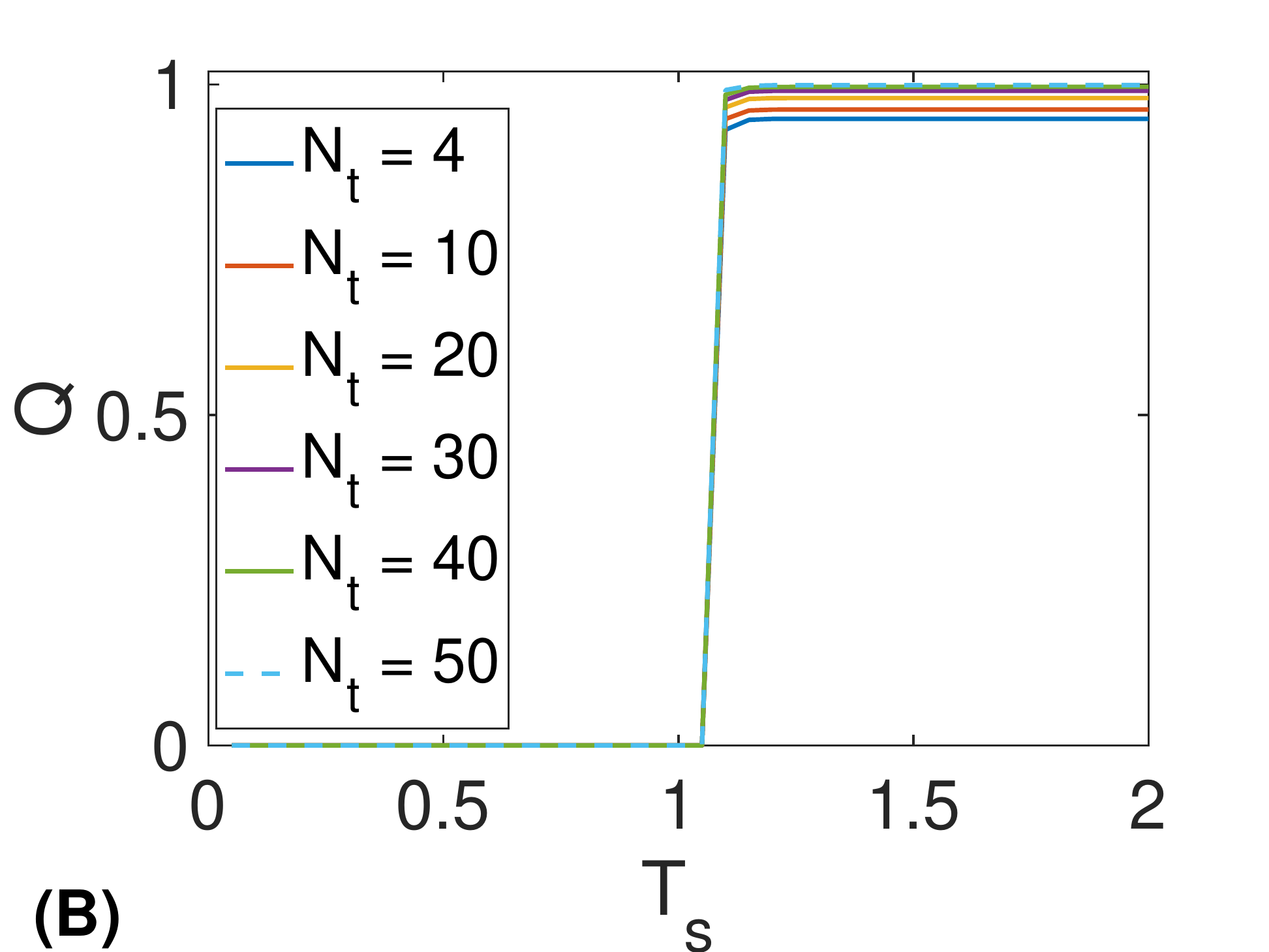} 
\includegraphics[scale=0.215]{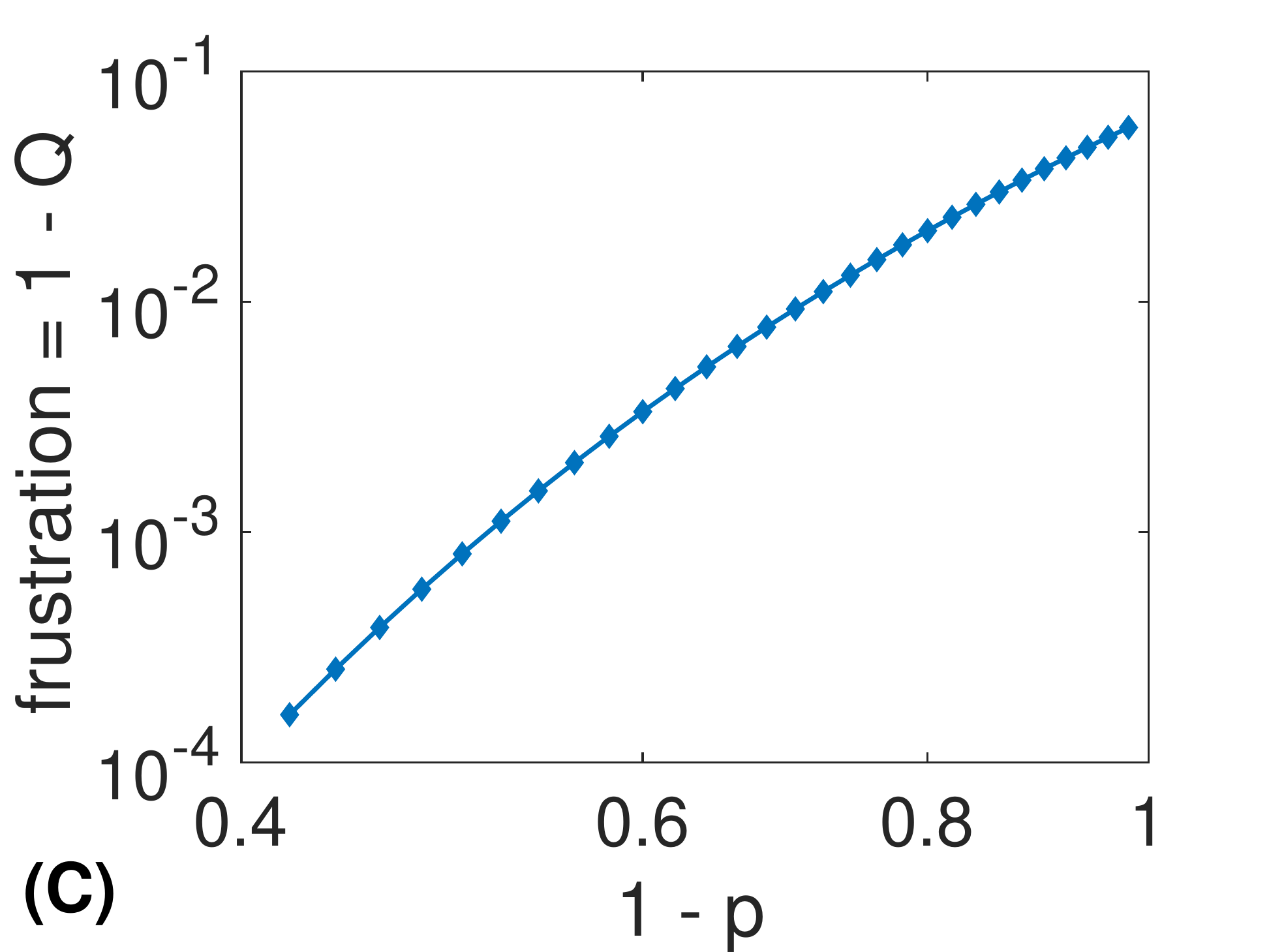}
 \includegraphics[scale=0.215]{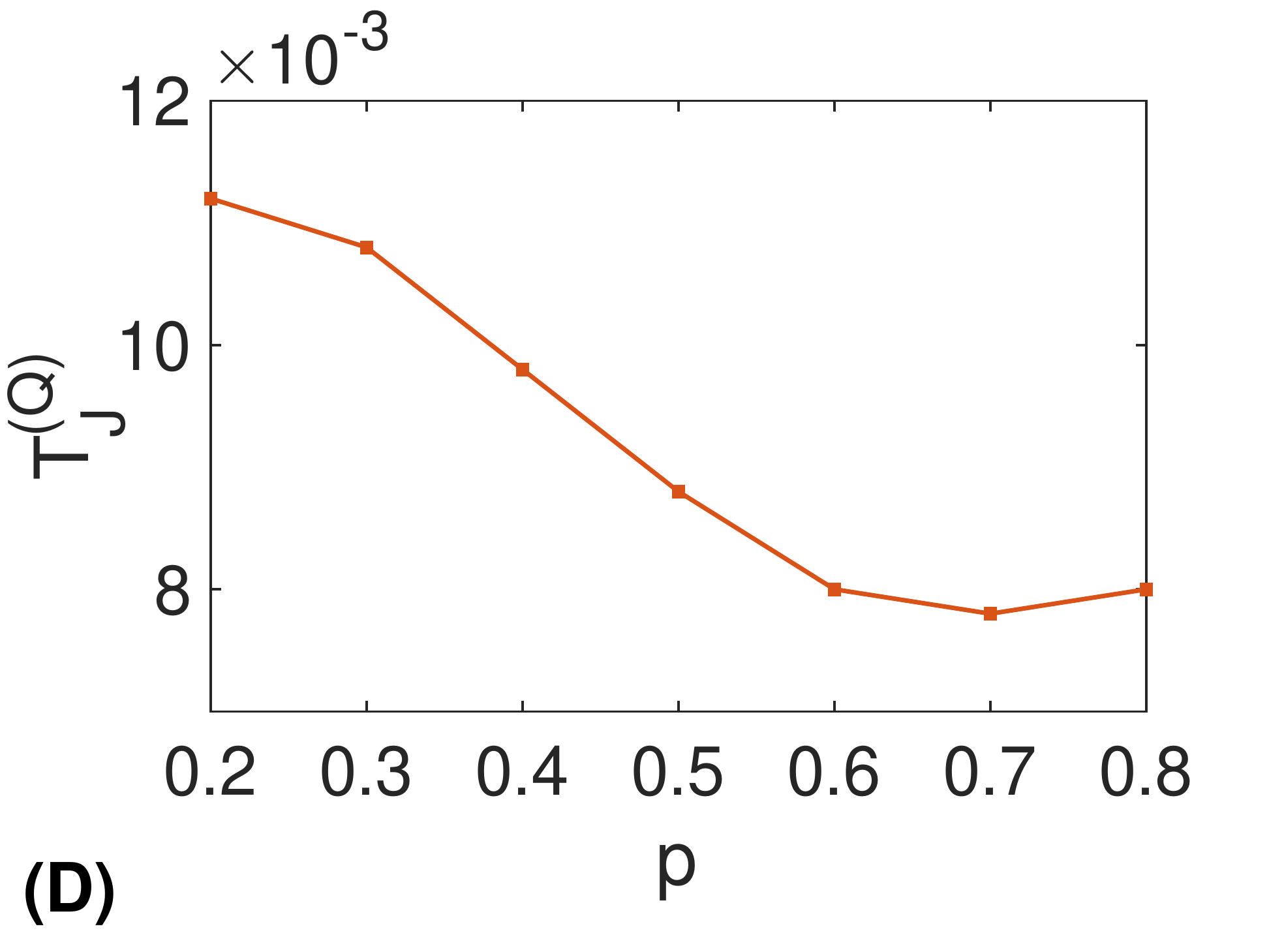}
  \includegraphics[scale=0.215]{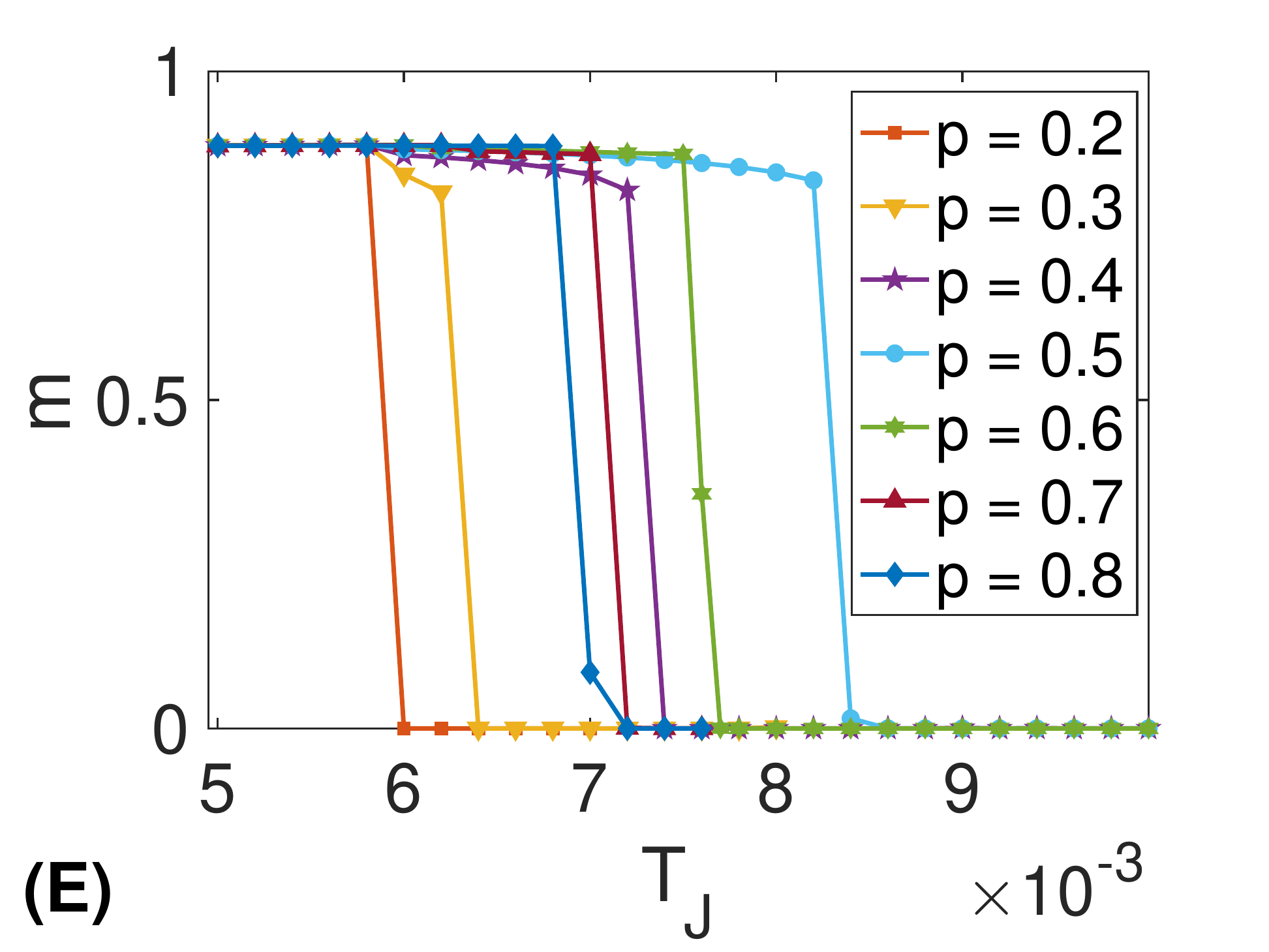}  \includegraphics[scale=0.215]{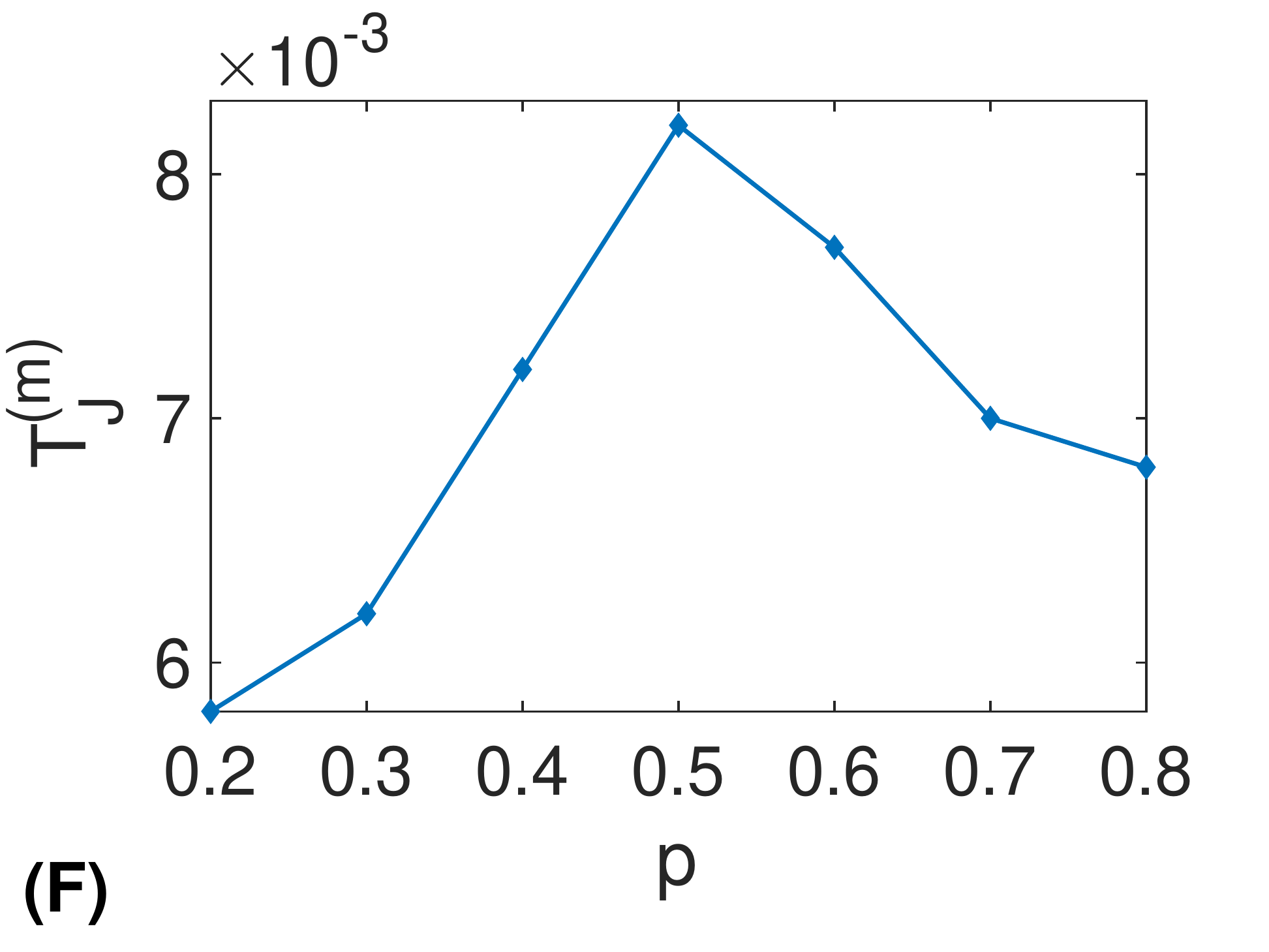}
\caption{ \textbf{(A)} Magnetisation of target spins $m$  as function of $T_s$ at fixed  $T_J = 0.005$ for various number of target spins \linebreak  $p = 0.04,0.1,0.2,0.3,0.4,0.5$.  \textbf{(B)} The same for genetic overlap $Q$.  \textbf{(C)} Frustration defined as $1 - Q$ as function of the number of non-target spins in the robust fitted phase (i.e. $T_s \in [T_c^{(1)}, T_c^{(2)}]$) at fixed  $T_J = 0.005$. \textbf{(D)} The highest value of $T_J$ at which $Q$ remains non-zero as function of $p$. \textbf{(E)}  Magnetisation of target spins $m$ as function of $T_J$ at fixed $T_s = 1.3$ for various fraction of target spins  $p = 0.2,0.3,0.4,0.5, 0.6, 0.7, 0.8$.  \textbf{(F)} The value of $T_J$ at which the magnetisation of target spins $m$ drops to zero at fixed $T_s = 1.3$ for different $p$ from \textbf{(E)}. Similar behaviour to \textbf{(E)} and \textbf{(F)}  is observed for others $T_s \in [T_c^{(1)}, T_c^{(2)}]$. Here  $N = 100$.}  
\label{fig:fig4}
\end{figure}

 The most relevant region in the phase diagram is robust fitted phase  \textbf{R}, which is characterized by both the high fitness ($m >0$) and robustness ($Q>0$). For a sufficiently low given $T_J$ (i.e. high selection pressure), the phase is bounded by $T_s \in [T_c^{(1)}, T_c^{(2)}]$. Below $T_c^{(1)}$, $Q$ goes to zero, and above $T_c^{(2)}$, $m$ goes to zero, whereas these transition points depend on $T_J$.  We first investigate the dependence on $p$ of the  \textbf{R} region by fixing $T_J$.
In Fig.  \ref{fig:fig4} \textbf{(A)} and \textbf{(B)} we fixed  $T_J = 0.005$.
First, for a wide range of $p \in [0.04, 0.5]$, the temperature $T_c^{(1)}$ of the latter transition in \textbf{(B)} does not depend on $p$ (see SM for the zoom-in of a small dip of $m$ at this point).
On the other hand, $T_c^{(2)}$ slightly increases with increasing $p$ in \textbf{(A)}, indicating that  the fitness of the high $p$ case is relatively more robust to noise than the low $p$ case.
%{\re ** However for higher $T_J$, this increase …?**}
 
In Fig. \ref{fig:fig4}  \textbf{(C)} $1-Q$ is almost constant against $T_s$ in the  \textbf{R} phase. This constant value was found to increase with the number of non-target spins.  Note that $Q\sim 1$ implies that the offspring of genotypes are preserved. The increase in $1-Q$, thus means the increase in redundancy of genotypes, as is supported by a larger number of non-targets.
Such a redundancy has another meaning in the context of spin-glass systems, where it is indeed equal to the local frustration (in the $(t-o-t)$ triads with positive $J^{(tt)}$).
 
In contrast, apart from the two critical points $T_c^{(1)}$ and $T_c^{(2)}$, the fitness $m$ does not depend on $p$. It follows a unique curve, independent of $p$.
Even though the increase in genetic heterogeneity $1-Q$ for more non-targets may perturb the target spin configuration, the fitness $m$ remains unchanged even for smaller $p$. %{\re Though this result has been obtained from a thermodynamic limit consideration, it  agrees  with the  numerical studies for finite $N$  \cite{Sakata2009}}.
 
Then, we estimate the critical value of $T_J$ below which the  \textbf{R} phase can exist. While this critical value denoted by $T_J^{(Q)}$ can depend on $T_s$, as see in Fig \ref{fig:fig2} and \ref{fig:fig3},  it  can be approximately identified with the upper part of the \textbf{P2} phase  from the phase diagram.  In Fig.  \ref{fig:fig4}   \textbf{(D)} $T_J^{(Q)}$ is shown to decrease with $p$. This result together with that in  Fig.  \ref{fig:fig4}   \textbf{(A)} mean that the higher $p$ is, the higher selection pressure is needed to achieve robustness, but once it is achieved,  a system with larger $p$ is  more robust with respect to phenotypic noise than  one with smaller $p$.
 
Finally, we examine the fitness as function of $T_J$ at fixed $T_s = 1.3$ for various $p$ in Fig. \ref{fig:fig4}   \textbf{(E)}. While fitness  decreases with $T_J$, its behaviour with the  increase in $p$ is non-monotonic. This behavior is further shown in Fig. \ref{fig:fig4}   \textbf{(F)} where the critical genotypic noise $T_J^{(m)}$ at which the fitness becomes non-zero is plotted versus $p$. Similar behavior is observed for others $T_s \in [T_c^{(1)}, T_c^{(2)}]$. The result supports $p=0.5$ as the maximal value of $T_J^{(m)}$, suggesting the existence of an optimal fraction of target spins to  acquire high fitness in this intermediate range of $T_s$ \footnote{This optimality, however, does not happen in the spinglass phase \textbf{SP2} with low temperature $T_s$, since the selection pressure needed to achieve non-zero fitness crucially depends on $p$. In fact as $p$ is increased, higher pressure (i.e. lower $T_J$) is needed to achieve non-zero $m$ of targets, i.e, to transition from \textbf{SP1} to \textbf{SP2}.}. 

 \section{Mutational susceptibility and Phenotypic susceptibility in the robust fitted phase}

Correlation between variances of phenotypes due to genetic changes and to noise has been discussed both in experiments and simulations, and relationships to robustness have been discussed both theoretically \cite{Sato2003,Kaneko2006, KanekoPloSOne2007,Ciliberti, Sakata2020, Tang2021} and experimentally \cite{Landry, Felix,  Silva-Rocha, Uchida}.
In statistical physics, this issue can be analyzed in terms of susceptibility, as it is proportional to the variance. Then, we need to study the susceptibility due to genetic mutation, in addition to the standard susceptibility.

In the context of this model, mutations are defined as those change of the genotypes $J_{ij}$ that might happen spontaneously and independently from the dynamics specified previously.  Let $\delta \Psi_i(\delta J_{jk})$ denote the change of the average local magnetisation of a target spin $i$ \footnote{More precisely, in \cite{Sakata2020} $\Psi_i = \langle {\rm sign}\,(m) s_i\rangle_{T_s}$.} upon mutating a genotype  $J_{jk} \rightarrow J_{jk} + \delta J_{jk}$. The mutational susceptibility of this target spin w.r.t such a change $M_{i,jk}$ then can be defined as $$M_{i,jk} = \lim_{\delta J_{jk} \rightarrow 0} \delta \Psi_i(\delta J_{jk})/ \delta J_{jk} = \big\langle s_is_js_k\big\rangle - \big\langle s_i\big\rangle  \big\langle s_js_k\big\rangle\,.$$
In general, $J_{jk} \in\bold{J} = \bold{J}^{(tt)} \cup \bold{J}^{(oo)} \cup \bold{J}^{(to)}$. However, since fitness is determined solely by the configurations of target spins at equilibrium, we consider only  $J_{jk} \in \bold{J}^{(tt)}$ and  show that %{\ re WE NEED FIRST INTRODUCE mutational sensitiviety, and then represent it be eq(12)} the  so-called 
the average of this mutational susceptibility %$  M_{i,jk}  := \big\langle s_is_js_k\big\rangle - \big\langle s_i\big\rangle  \big\langle s_js_k\big\rangle $, after averaging 
over all triples  $(i,j,k) \in \mathcal{T}$ is equal to 
 \begin{equation}
     M := \frac{1}{\binom{N_t}{3}} \sum_{(i,j,k)}M_{i,jk}  = 2\beta_J^2 m \chi_m - \beta_J \frac{\partial^3 f}{\partial h^3}\Big|_{h = 0} \\
%& =\beta_s^3 \left[2m n \,\frac{\sum_{a,b}^n {\rm Tr} \,s^as^b e^L}{{\rm Tr}\, e^L} + \frac{\sum_{a,b,c}^n {\rm Tr}\, s^as^bs^c e^L}{{\rm Tr}\, e^L} \right]
\label{mutational_sus}
 \end{equation}
 where  $\chi_{m} :=  - \lim_{h\rightarrow 0}  \partial^2 f/\partial h^2$  is the susceptibility  of target  spins.  The quantities $M$ and $\chi_m$ correspond to the susceptibility to mutation of the genotypes and susceptibility to perturbation by an external field, $h$, respectively. 
We can expect that  in the robust fitted phase there exists a relation between $M$ and $\chi_m$ \footnote{From the replica-symmetric free energy density $f_{\mathcal{T}}^{\rm RS}$ we get $\chi^{\rm RS}_{m} = \beta_s \big[1 + (n-1) q\big]$, while $M$ can not be obtained directly within this ansatz.}. In fact, let $X := \lim_{h\rightarrow 0} \partial^3 f_{\mathcal{T}}/ \partial h^3$. %
       For $L$  given in the SM Eq. \eqref{weight_target} ($L/\beta_J$ has the meaning of an effective Hamiltonian that is defined in the combined space $\{s^a_i, \sigma^k_i\}$ of $s$-replicas and $J$-replicas), according to the definitions
 \begin{eqnarray*}
 X \propto \sum_{a,b,c=1}^n {\rm Tr} \,(s^as^b s^c e^L)/{\rm Tr} \,(e^L) \,,
 \\
 m := \langle s^a \rangle=  \frac{1}{n}\,\sum_{a=1}^n {\rm Tr} \,(s^ae^L) /{\rm Tr} \,(e^L) \,,
 \\\chi_{m} :=  \lim_{h\rightarrow 0}\frac{\partial m}{ \partial h} = \frac{\beta_s}{n}\, \sum_{a,b=1}^n {\rm Tr} \,(s^as^be^L)/{\rm Tr} \,(e^L) \,,
 \end{eqnarray*}
the  symmetry between different replicas in the robust fitted phase implies that the third moment $X^{\rm RS}$ is proportional to the product of the first and second moments 
$m^{\rm RS}\chi_m^{\rm RS}$. %$$= \frac{{\rm Tr} \,(s\cdot e^L)} {{\rm Tr} \,(e^L)} \sum_{b,c=1}^n \frac{{\rm Tr} \,(s^bs^ce^L)}{{\rm Tr} \,(e^L)}  \propto X^{\rm RS}\,, $$
 Therefore,   approximately, $M \propto \chi_m^{\rm RS}$.   This proportionality between the two susceptibilities, implying a correlation between phenotypic changes due to genetic variation  and those in response to environmental perturbations \cite{Lehner2010}, does not exist in the RSB phase, as the second term in Eq. \eqref{mutational_sus} is no longer proportional to $\chi_m$.  
 %As argued in \cite{Sakata2020} such correlation can only be achieved in the replica symmetric region where the original high-dimensional dynamics of the phenotypes is reduced to a low-dimensional manifold due to evolution towards robustness. The variation of fitness due to noise and that due to mutation then happen to  occur along the same low-dimensional manifold, resulting in a correlation between them. {\re If RSB occurs, such restriction of  the  phenotypic dynamics no longer  exits, because in this case, changes of fitness upon varying the environmental conditions will vary arbitrarily from realisation to realisation of the $J_{ij}$'s dynamics.  As a result,  the system will have random, uncorrelated responses to noise and to mutations.}
 
 \section{Discussion}
 In the paper we propose a new approach towards  biological evolution due to the interrelationship  between genotype and phenotype where fitness is determined solely by the latter but not by the former. Though the emergence of structured  genotypes  from initially random  couplings under this relation has been  numerically reported, apart from a study which imposed a specific condition on the couplings \cite{Sakata2012}, this has not been studied analytically yet. We here are able to tackle this problem thanks to what we termed \emph{double-replica} theory. Within this framework we  obtain the  phase diagram,  that is classified not only by the fitness but also by the overlap in dual replicas. The diagram is not only in good agreement with previous studies (including paramagnet, t-ferro and robust fitted phases, all existing at sufficiently low $T_J$), but also contains previously undiscovered phases. These include the first spin-glass phase \textbf{SP1} and the second paramagnet phase \textbf{P2}. The former corresponds to a system with both target and non-target spins residing in a spin-glass phase (at low selection pressure),  while the latter corresponds to a paramagnetic phase for all spins but with retaining genetic correlations  encoding target- and non-target couplings (at high  selection pressure and high $T_s$).  Here even though the genotypes favor a high value of fitness, due
to large fluctuations induced by $T_s$, such value can not be maintained over generations. { The existence of the phase suggests that even though the average fitness is zero due to large noise, there exists genetic precursor to generate individuals with non-zero fitness.  The relevance of this scenario  to evolutionary biology, needs to be explored in future, though.}

 The system can  only acquire high fitness at some $T_s \leq T_c^{(2)}$, where the fitness increases discontinuously. If $T_s$ is too low, then  RSB will happen, to a phase without \emph{genetic} overlap, where biologically required robustness of genotypes is lost.  Hence a lower bound of $T_s \geq T_c^{(1)}$ is necessary to have RS and  robustness, accordingly. %{\re in some of these, the high fitness can exist at high selection pressure}
 
 From this approach, the target-fraction dependence of genotypic and phenotypic robustness can also be understood quantitatively. Such dependence is quantified via the behaviour of the fitness $m$ and  genetic redundancy $1-Q$ in the robust region bounded by $T_c^{(1)}$ and $T_c^{2}$. Here we find that 
 %was further shown to be tightly-related to a trade-off between fitness and genetic redundancy: 
 a genetically homogeneous population can only be robustly reproduced under a sufficiently high selection pressure  and under a sufficient level of  phenotypic noise (temperature).   As the fraction of target spins is increased, the robust fitted phase is slightly expanded to a higher temperature, whereas higher selection pressure is needed to achieve it.  The existence of an optimal fraction for attaining high fitness under  intermediate phenotypic noise is suggested.  This may explain why, in biological systems, such as in proteins, the fraction of units that are responsible for function is generally limited, and a sufficient fraction of non-functional units is needed, providing redundancy.
 
  In the present theory, the proportionality between the standard thermodynamic susceptibility and mutational susceptibility is derived in the robust fitted phase. As the susceptibility  measures  the change of fitness due to varying conditions, a correlation between responses to  environmental perturbations and that by genetic changes is suggested. Such correlation, or evolutionary fluctuation-response relationship \cite{KanekoPloSOne2007, Sakata2020, Sato2003, Kaneko2006, Ciliberti, Tang2021} has been observed in  experimental data from the evolution of protein dynamics and bacterial protein expressions, whereas we can derive it here under replica symmetry assumption. As argued in \cite{Sakata2020}, such correlation can only be achieved in the replica symmetric region where the original high-dimensional dynamics of the phenotypes are reduced to a low-dimensional manifold due to evolution towards robustness. The variation of fitness due to noise and that due to mutation then happen to  occur along the same low-dimensional manifold, resulting in a correlation between them. If RSB occurs, such restriction of  the  phenotypic dynamics no longer  exits, because in this case, changes of fitness upon varying the environmental conditions will vary arbitrarily from realisation to realisation of the $J_{ij}$'s dynamics.  As a result,  the system will have random, uncorrelated responses to noise and to mutations.

 In our formulation by assuming that the entire system  reaches an equilibrium, we approximate the effect of the slowly-evolving $\bold{J}^{(to)}$ that couple the subsystem $S_{\mathcal{T}}$ to $S_{\mathcal{O}}$ on these subsystems' own dynamics by the equilibrium correlations $\langle J^{(to)} J^{(to)} \rangle$. Such correlations are then incorporated separately into each of the dynamics of the $\bold{J}^{(tt)}$ and $\bold{J}^{(oo)}$ couplings by modifying the effective potentials of these dynamics, thus making the dynamics of these two different sets of coupling \emph{independent} of each other. In an equilibrium statistical physics formulation, this leads to the necessity of introducing  another type of replica, so-called coupling-replicas into  the partition functions, besides the first (standard) spin-replicas that take care of the effect of the phenotypes on the evolution of genotypes.  This scheme hence allows us to treat the model in a standard mean-field manner. On one hand, being of mean-field nature, our approach can not provide a formal argument to support the hypothesis of \cite{Sakata2020} about the emergent dimensional reduction from phenotype-genotype co-evolution. 
 On the other hand, the correlation between the mutational- and environmental susceptibility in the robust fitted phase suggests the existence of  a funnel-like landscape \cite{Go1983, Onuchic} that reinforces the dynamics to reside in a low-dimensional manifold by a global attraction. 
 
 The current choice of fitness for the sake of simplicity, however, limits the possibility of having different global maxima in the fitness landscape. One can enrich the model behavior by  determining  fitness either by a combination of $N_{\rm fit}$ different target spin configurations or by a set of gauge-equivalent configurations. %Furthermore, since the current fitness function does not have any explicit dependence on non-targets, the targets can become ferromagnetic themselves without the aid of non-targets. So a natural question is how to generalise the definition of fitness to take such an effect into account. A recently proposed reciprocity relation between target and non-target \cite{Hatakeyama2020} may be a good candidate for this task. On another note, the model \cite{Sakata2009} only deals with a predefined subset of target spins, or more precisely, their fraction without considering the   spontaneous emergence of this subset from  a population of unassigned (undiscriminated) spins. It would be possible to generalise the model by incorporating  such a dynamics of the network that gives rise to the structure  with a \emph{core}  identified as the subset of target spins.
 
 In the present framework, since the couplings are symmetric,  we  constructed the effective potential of the coupling dynamics  based partly on the existence of an energy landscape. For those models in other contexts \cite{Poderoso2007, Pham2020} having such a landscape picture, we expect a  straightforward application of our  approach. Furthermore, the present double-replica theory can  be extended to those stochastic dynamical systems that are not governed by Hamiltonian dynamics as well.   In this case, instead of the effective potential and its associated partition function, one would need to charaterize the ensemble of trajectories in the combined space of phenotypes and genotypes, using the moment generating function \cite{Martin, Dominicis}. While we so far have solely used  phenotypes and  genotypes as the main example of our approach, such an extension
would allow for the applications to co-evolution of gene-expression patterns and the gene-regulatory networks \cite{KanekoPloSOne2007}, that of species abundances and their ecological networks \cite{Barbier}, and that of neuronal activities and  network shaped by neural dynamics (learning) \cite{Kadmon2015,Schuecker2018}.
%Mastrogiuseppe2017 % landscape  such as gene expression dynamics \cite{KanekoPloSOne2007} or neural networks with asymmetric couplings \cite{Schuecker2018, Mastrogiuseppe2017}. 
\bigskip

%https://www.overleaf.com/project/6353b6dc7c2bad368e9c3b72
\begin{acknowledgments}
We acknowledge support from Novo Nordisk Foundation and would like to thank Ayaka Sakata, Koji Hukushima and Yoshiyuki Kabashima for stimulating discussion.
\end{acknowledgments}

\bibliography{Doublereplica}% Produces the bibliography via BibTeX.
 
\clearpage
 \appendix
 
 \onecolumngrid
 \section*{Supplemental material}
 \subsection*{A. Details  of the SHK model}
In the following we call the model originally introduced in \cite{Sakata2009, Sakata2009PRE} as the SHK model. In this model,  phenotypes are spin configurations, and genotypes are the interaction matrix for spins. In a system of $N$ spins,  each spin $i$ can take values $s_i \in \{-1,1\}$ and is linked to exactly $N-1$ other spins, thus forming a fully-connected network.  Moreover, fitness is determined by a subset of  target spins denoted by  $\mathcal{T}$.  Those spins that do not contribute to the fitness are called non-target. The fitness $\Psi$ at a noise level $T_s$  is determined by the spin configurations at equilibrium as  
  \begin{equation}
  \Psi(\bold{s}) = \frac{1}{N_t}\,\left \langle  \Big|\sum_{i \in \mathcal{T}} s_i\Big| \right\rangle_{T_s}  \,, \label{fitness_SM}
  \end{equation}
  where $\langle\cdot \rangle_{T_s}$ is the thermal average according to an equilibrium distribution over spin configurations only. Such distribution is computed from the partition function of a  spin-glass Hamiltonian  $H_S = -\sum_{i<j} J_{ij}s_i s_j$ \cite{Sherrington} in which the couplings $J_{ij}$ are regarded as \emph{fixed} over the course of the spin dynamics because they  are assumed to evolve on much slower timescale than that of the spins. Here the couplings are  symmetric, i.e.  $J_{ij} = J_{ji}$, and are  independently and identically distributed by a Gaussian distribution with  zero mean and the variance \linebreak $J^2 := {\rm var}(J_{ij}) = N^{-1}$.  The  model Hamiltonian of the full system  is given by
\begin{equation}H_{S} = - \sum_{i<j} J_{ij} s_is_j \label{original_spin_hamiltonian} \end{equation} 
 Once the spins have relaxed to an equilibrium at a temperature  $T_s$ via a Glauber update specified by $H_S$,  the couplings are next updated  with probability ${\rm Pr}\big[\bold{J} \rightarrow \bold{\tilde{J}}\big] = {\rm min}\,\big\{1,e^{\beta_J \Delta\Psi}\big\}$, where $\Delta\Psi = \Psi(\bold{\tilde{J}}) - \Psi(\bold{J})$ and $\beta_J\equiv 1/T_J$ is the genotypic selection pressure. These two dynamics are implemented consecutively one after another until the entire system equilibrates. Implementing this way, the model captures the  evolution of feedback process between the phenotype and genotype, where the phenotype dynamics are represented by the stochastic dynamics of spins $(\bold{s})$ according to the energy landscape $H_S$ for given genotype $(\bold{J})$,  whereas the evolution of genotype  is given by the stochastic change of $(\bold{J})$  according to the fitness $ \Psi(\bold{s})$ determined by the phenotype. On the contrary to more common theories of evolution, this model hence explicitly considers the co-evolution of these coupled landscapes.

%Up to the  order  $1/N_t$, the average frustration among target spins 
 %\begin{align} \nonumber    \binom{N-N_t}{3}\cdot \Big\langle J_{ij}^{(oo)} J_{jk}^{(oo)} J_{ki}^{(oo)}\Big\rangle &= 3 \tilde{M}^2(\tilde{q}+Q) + \frac{\big(1+(n-1)\tilde{q}\big)^3 +(n-1)(1-\tilde{q})^3}{n^3} \\ &+ \frac{\big(1+(t-1)Q\big)^3 +(t-1)(1-Q)^3}{t^3}   +O\left(\frac{1}{t}\right)  \end{align} 
    \subsection*{B. Replica symmetric ansatz solution and the expression of $I_{kz}$ and $\tilde{I}_k$}
      The partition functions  are given in terms of the  \emph{target} and  the \emph{non-target}  free energy densities, $f_{\mathcal{T}}(\bold{m},\bold{q},\bold{r},\bold{Q},\bold{M})$ and $   f_{\mathcal{O}}(\bold{\tilde{m}},\bold{\tilde{q}},\bold{\tilde{r}},\bold{\tilde{Q}},\bold{\tilde{M}})$, respectively,  by
       \begin{subequations}
\label{allequations9}
 \begin{eqnarray}
\mathcal{Z}_{\mathcal{T}}  = \int D\bold{m}\, D\bold{q}\,
   D\bold{r}\,D\bold{Q}\, D\bold{M} \,e^{-\beta_J Np f_{\mathcal{T}}(\bold{m},\bold{q},\bold{r},\bold{Q},\bold{M})}
  \label{partition1}
\\
   \mathcal{Z}_{\mathcal{O}}      = \int D\bold{\tilde{m}}\, D\bold{\tilde{q}}\,
   D\bold{\tilde{r}}\,D\bold{\tilde{Q}}\, D\bold{\tilde{M}} \,e^{-\beta_J N(1-p) f_{\mathcal{O}}(\bold{\tilde{m}},\bold{\tilde{q}},\bold{\tilde{r}},\bold{\tilde{Q}},\bold{\tilde{M}})}\label{partition2}
\end{eqnarray}
\end{subequations}
  where
         \begin{subequations}
\label{allequations10}
 \begin{eqnarray}
 f_{\mathcal{T}} =  \frac{ 1}{2}   \,\Big\{ \sum_{a<b} \frac{q_{ab}^2}{n^2} + \sum_{k<k'} \frac{Q_{k,k'}^2}{(N-t)^2} + \frac{1}{n(N-t)}\sum_{a,k} M_{ak}^2\Big\}  -\frac{1}{\beta_J} \ln \sum_{\big\{s_i;s^\alpha;\sigma^k\big\}_{i\in \mathcal{T}}} e^L
   \label{free_energy_target}
\\
  f_{\mathcal{O}} =  \frac{1}{2}   \,\Big\{\tilde{K} \sum_a \frac{\tilde{m}^2_a}{n} + \tilde{K} \sum_k \frac{\tilde{r}^2_k}{t} +   \sum_{a<b} \frac{\tilde{q}_{ab}^2}{n^2} + \sum_{k<k'} \frac{\tilde{Q}_{k,k'}^2}{t^2} + \frac{1}{nt}\sum_{a,k} \tilde{M}_{ak}^2\Big\}  - \frac{1}{\beta_J}  \ln \sum_{\big\{s^\alpha;\sigma^k\big\}} e^{\tilde{L}}
    \label{free_energy_non_target}
\end{eqnarray}
\end{subequations}
         \begin{subequations}
\label{allequations11}
 \begin{eqnarray}
\nonumber L&=& \frac{\beta_J}{4N_t}\, \Big|\sum_{i \in \mathcal{T}} s_i\Big|^2 -\frac{\beta_J}{2}\, \Big|\sum_{i \in \mathcal{T}} s_i\Big| \left[\sum_{a=1}^{n} \frac{m^2_a}{n} + \sum_k^{N-N_t} \frac{r^2_k}{N-N_t}\right] \\ &+&\beta_J \left(\frac{1}{N_t}\, \Big|\sum_{i \in \mathcal{T}} s_i\Big| \left[\sum_{a=1}^n \frac{m_a s^a}{n} + \sum_k^{N-N_t} \frac{r_k \sigma^k}{N-N_t}\right] +\sum_{a<b}^n\, \frac{ q_{ab} s^a s^b}{n^2}+  \sum_{k<k'}^{N-N_t}\,\frac{Q_{kk'} \sigma^k\sigma^{k'}}{(N-N_t)^2}\,  +  \sum_{a,k}\, \frac{M_{ak} s^a \sigma^k}{n(N-N_t)} \right)
 \label{weight_target}
\\
 \tilde{L} &=& \beta_J \left(\frac{\tilde{K} }{n}\sum_{a=1}^n \tilde{m}_a s^a + \frac{\tilde{K} }{N_t}\sum_k^{N_t} \tilde{r}_k \sigma^k + \frac{1}{n^2}\, \sum_{a<b}^n\, \tilde{q}_{ab} s^a s^b+ \frac{1}{N_t^2}\, \sum_{k<k'}^{N_t}\, \tilde{Q}_{kk'} \sigma^k\sigma^{k'} + \frac{1}{n N_t}\, \sum_{a,k}\, \tilde{M}_{ak} s^a \sigma^k  \right)
    \label{weight_non_target}
\end{eqnarray}
\end{subequations}
  Denoting $\mathcal{D}x \mathcal{D}y= \frac{e^{-(x^2+y^2)/2}}{2\pi}\,dx dy$ and $A_z(m,r) =   \frac{\beta_J}{4}\frac{(N_t-2z)^2}{N_t^2} - \frac{\beta_J}{2}\frac{|N_t-2z|}{N_t}\big(m^2 +r^2\big)$ we have
    
      \begin{subequations}
\label{allequations7}
 \begin{eqnarray}
\nonumber I_{k,z} &=& \int \mathcal{D}x \mathcal{D}y\, \exp\left\{A_z + \frac{N-N_t-2k}{N-N_t}\,\Big(y\sqrt{\beta_J Q} + \beta_J r\frac{|1-2z|}{N_t}  \Big)\right\} \\ 
&\times& \left[ {\rm cosh}\Big(\beta_s m\, \frac{|N_t-2z|}{N_t}   + x \frac{\sqrt{\beta_J q}}{n} +\frac{\beta_J M}{n}\, \frac{N-N_t-2k}{N-N_t}\Big) \right]^n
  \label{integral1}
\\
 \tilde{I}_{k} &=& \int \mathcal{D}x \mathcal{D}y\, \exp\left\{\frac{N_t-2k}{N_t}\,\Big(y\sqrt{\beta_J  \tilde{Q}} + \beta_J \tilde{K}  \tilde{r} \Big)\right\}\,\left[ {\rm cosh}\Big(\beta_s \tilde{K}  \tilde{m} + x \frac{\sqrt{\beta_J  \tilde{q}}}{n} +\frac{\beta_J  \tilde{M}}{n}\, \frac{N_t-2k}{N_t}\Big) \right]^n
  \label{integral2}
\end{eqnarray}
\end{subequations}
The argument of the cosh$(\cdot)$ function will be denoted by
       \begin{equation}\Omega = \beta_s m\, \frac{|N_t-2z|}{N_t}   + x \frac{\sqrt{\beta_J q}}{n} +\frac{\beta_J M}{n}\, \frac{N-N_t-2k}{N-N_t}\,,\qquad \tilde{\Omega} = \beta_s \tilde{K}  \tilde{m} + x \frac{\sqrt{\beta_J  \tilde{q}}}{n} +\frac{\beta_J  \tilde{M}}{n}\, \frac{N_t-2k}{N_t} 
       \label{omega}
       \end{equation}
The replica symmetric free energy densities given in the main text yield the extremum condition after setting $\tilde{K} = 0$ 
%\label{allequations}

 \begin{eqnarray*}
m & = &\frac{\displaystyle \sum_{z=0}^{N_t} \binom{N_t}{z}    \sum_{k=0}^{N-N_t} \binom{N-N_t}{k}  \int \mathcal{D}x \mathcal{D}y \exp\left\{A_z +\frac{N-N_t-2k}{N-N_t} \,\Big(y\sqrt{\beta_J Q} + \beta_J r \theta_z \Big)\right\} \,\big[ {\rm cosh}(\Omega) \big]^n {\rm tanh}(\Omega)}{ \displaystyle \sum_{z=0}^{N_t}  \binom{N_t}{z}   \sum_{k=0}^{N-N_t} \binom{N-N_t}{k} \int \mathcal{D}x \mathcal{D}y\, \exp\left\{A_z +\frac{N-N_t-2k}{N-N_t}\,\Big(y\sqrt{\beta_J Q} + \beta_J r \theta_z \Big) \right\}\,\big[ {\rm cosh}(\Omega(x,q,M)) \big]^n } 
\\
    q & =& \frac{\displaystyle \sum_{z=0}^{N_t} \binom{N_t}{z}   \sum_{k=0}^{N-N_t} \binom{N-N_t}{k} \int \mathcal{D}x \mathcal{D}y\, \exp\left\{A_z +\frac{N-N_t-2k}{N-N_t}\,\Big(y\sqrt{\beta_J Q} + \beta_J r \theta_z \Big)\right\}\,\big[ {\rm cosh}(\Omega) \big]^n  \big[ {\rm tanh}(\Omega) \big]^2}{  \displaystyle \sum_{z=0}^{N_t}\binom{N_t}{z}   \sum_{k=0}^{N-N_t} \binom{N-N_t}{k} \int \mathcal{D}x \mathcal{D}y\, \exp\left\{A_z+\frac{N-N_t-2k}{N-N_t}\,\Big(y\sqrt{\beta_J Q} + \beta_J r \theta_z \Big)\right\}\,\big[ {\rm cosh}(\Omega) \big]^n  }  
    \\
   Q & = & -\frac{1}{N-N_t}+  \frac{\displaystyle\sum_{z=0}^{N_t}  \binom{N_t}{z}    \sum_{k=0}^{N-N_t} \binom{N-N_t}{k} \left(\frac{N-N_t-2k}{N-N_t}\right)^2 I_{k,z}}{ \displaystyle \sum_{z=0}^{N_t}  \binom{N_t}{z}   \sum_{k=0}^{N-N_t} \binom{N-N_t}{k} I_{k,z} }\,,\qquad
        r  = \frac{\displaystyle \sum_{z=0}^{N_t}  \binom{N_t}{z}   \sum_{k=0}^{N-N_t} \binom{N-N_t}{k} \frac{N-N_t-2k}{N-N_t}\,I_{k,z} }{ \displaystyle\sum_{z=0}^{N_t} \binom{N_t}{z}    \sum_{k=0}^{N-N_t} \binom{N-N_t}{k} I_{k,z} }
        \\
            M &=& \frac{\displaystyle \sum_{z=0}^{N_t} \binom{N_t}{z}   \sum_{k=0}^{N-N_t} \binom{N-N_t}{k} \frac{N-N_t-2k}{N-N_t} \int \mathcal{D}x \mathcal{D}y \exp\left\{A_z+ \frac{N-N_t-2k}{N-N_t} \,\Big(y\sqrt{\beta_J Q} + \beta_J r \theta_z \Big)\right\} \,\big[ {\rm cosh}(\Omega\big]^n {\rm tanh}(\Omega)}{ \displaystyle \sum_{z=0}^{N_t}  \binom{N_t}{z}   \sum_{k=0}^{N-N_t} \binom{N-N_t}{k} \int \mathcal{D}x \mathcal{D}y\, \exp\left\{A_z +\frac{N-N_t-2k}{N-N_t}\,\Big(y\sqrt{\beta_J Q} + \beta_J r \theta_z \Big)\right\}\,\big[ {\rm cosh}(\Omega) \big]^n } 
             \end{eqnarray*}
             \begin{eqnarray*}
        \tilde{m} &=& \frac{\displaystyle \sum_{k=0}^{N_t} \binom{N_t}{k}  \int \mathcal{D}x \mathcal{D}y \exp\left\{\frac{N_t-2k}{N_t} \,y\sqrt{\beta_J Q}\right\} \,\big[ {\rm cosh}(\Omega(x, \tilde{m}, \tilde{q}, \tilde{M})) \big]^n {\rm tanh}(\Omega(x, \tilde{m}, \tilde{q}, \tilde{M}))}{ \displaystyle \sum_{k=0}^{N_t} \binom{N_t}{k} \int \mathcal{D}x \mathcal{D}y\, \exp\left\{\frac{N_t-2k}{N_t}\,y\sqrt{\beta_J Q}\right\}\,\big[ {\rm cosh}(\Omega(x, \tilde{m}, \tilde{q}, \tilde{M})) \big]^n } \\
   \tilde{q} &=& \frac{\displaystyle \sum_{k=0}^{N_t} \binom{N_t}{k} \int \mathcal{D}x \mathcal{D}y\, \exp\left\{\frac{N_t-2k}{N_t}\,y\sqrt{\beta_J Q}\right\}\,\big[ {\rm cosh}(\Omega(x, \tilde{m}, \tilde{q}, \tilde{M})) \big]^n  \big[ {\rm tanh}(\Omega(x, \tilde{m}, \tilde{q}, \tilde{M}))\big]^2}{  \displaystyle \sum_{k=0}^{N_t} \binom{N_t}{k} \int \mathcal{D}x \mathcal{D}y\, \exp\left\{\frac{N_t-2k}{N_t}\,y\sqrt{\beta_J Q}\right\}\,\big[ {\rm cosh}(\Omega(x, \tilde{m}, \tilde{q}, \tilde{M})) \big]^n  } 
   \\
    \tilde{Q} & = & -\frac{1}{N_t}+  \frac{\displaystyle    \sum_{k=0}^{N_t} \binom{N_t}{k} \left(\frac{N_t-2k}{N_t}\right)^2 \tilde{I}_{k}}{ \displaystyle   \sum_{k=0}^{N_t} \binom{N_t}{k} \tilde{I}_{k} }  \,,\qquad
        \tilde{r}  =\frac{\displaystyle  \sum_{k=0}^{N_t} \binom{N_t}{k} \frac{N_t-2k}{N_t}\,\tilde{I}_{k} }{ \displaystyle  \sum_{k=0}^{N_t} \binom{N_t}{k} \tilde{I}_{k} }
        \\
        \tilde{M} &=& \frac{\displaystyle \sum_{k=0}^{N_t} \binom{N_t}{k} \frac{N_t-2k}{N_t} \int \mathcal{D}x \mathcal{D}y \exp\left\{\frac{N_t-2k}{N_t} \,y\sqrt{\beta_J Q}\right\} \,\big[ {\rm cosh}(\Omega(x, \tilde{m}, \tilde{q}, \tilde{M}))\big]^n {\rm tanh}(\Omega(x, \tilde{m}, \tilde{q}, \tilde{M}))}{ \displaystyle \sum_{k=0}^{N_t} \binom{N_t}{k} \int \mathcal{D}x \mathcal{D}y\, \exp\left\{\frac{N_t-2k}{N_t}\,y\sqrt{\beta_J Q}\right\}\,\big[ {\rm cosh}(\Omega(x, \tilde{m}, \tilde{q}, \tilde{M})) \big]^n }
 \end{eqnarray*}
    
 \subsection*{C. Phase diagram of the order parameters of the non-target spins and that of the eigenvalue of the Hessian}
\begin{figure}[t]
 \centering
\includegraphics[scale=0.25]{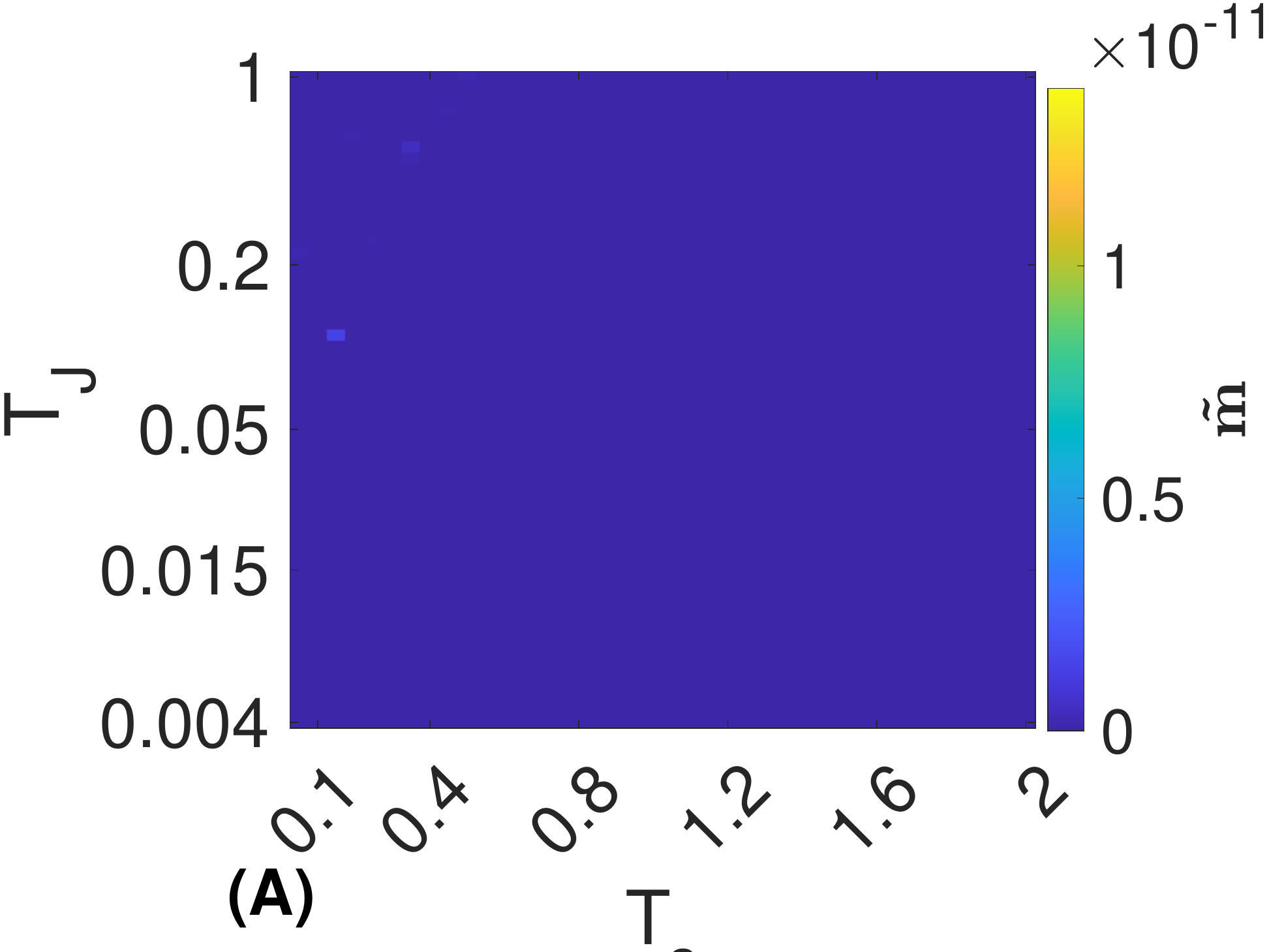}
\includegraphics[scale=0.25]{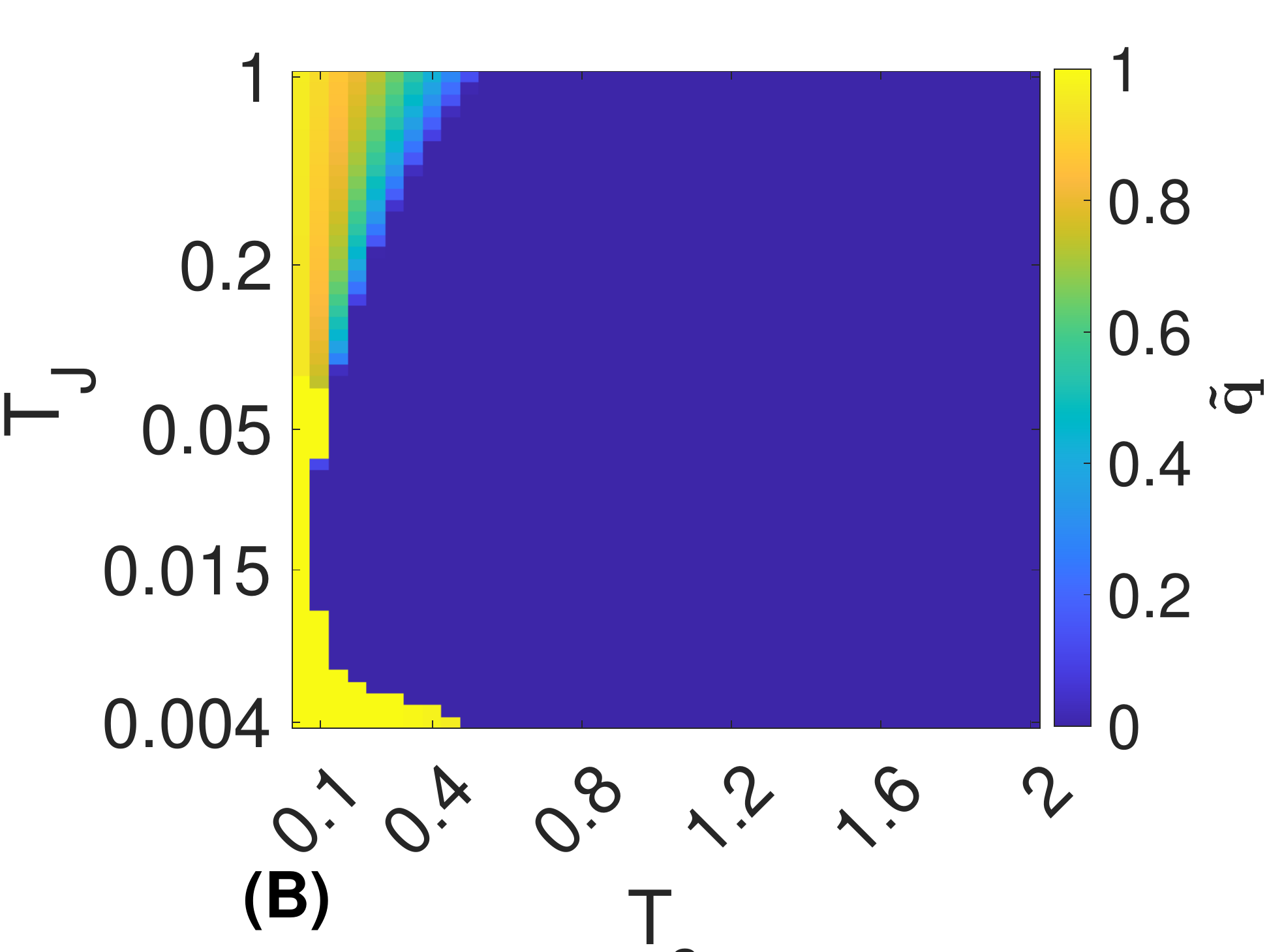}

\includegraphics[scale=0.25]{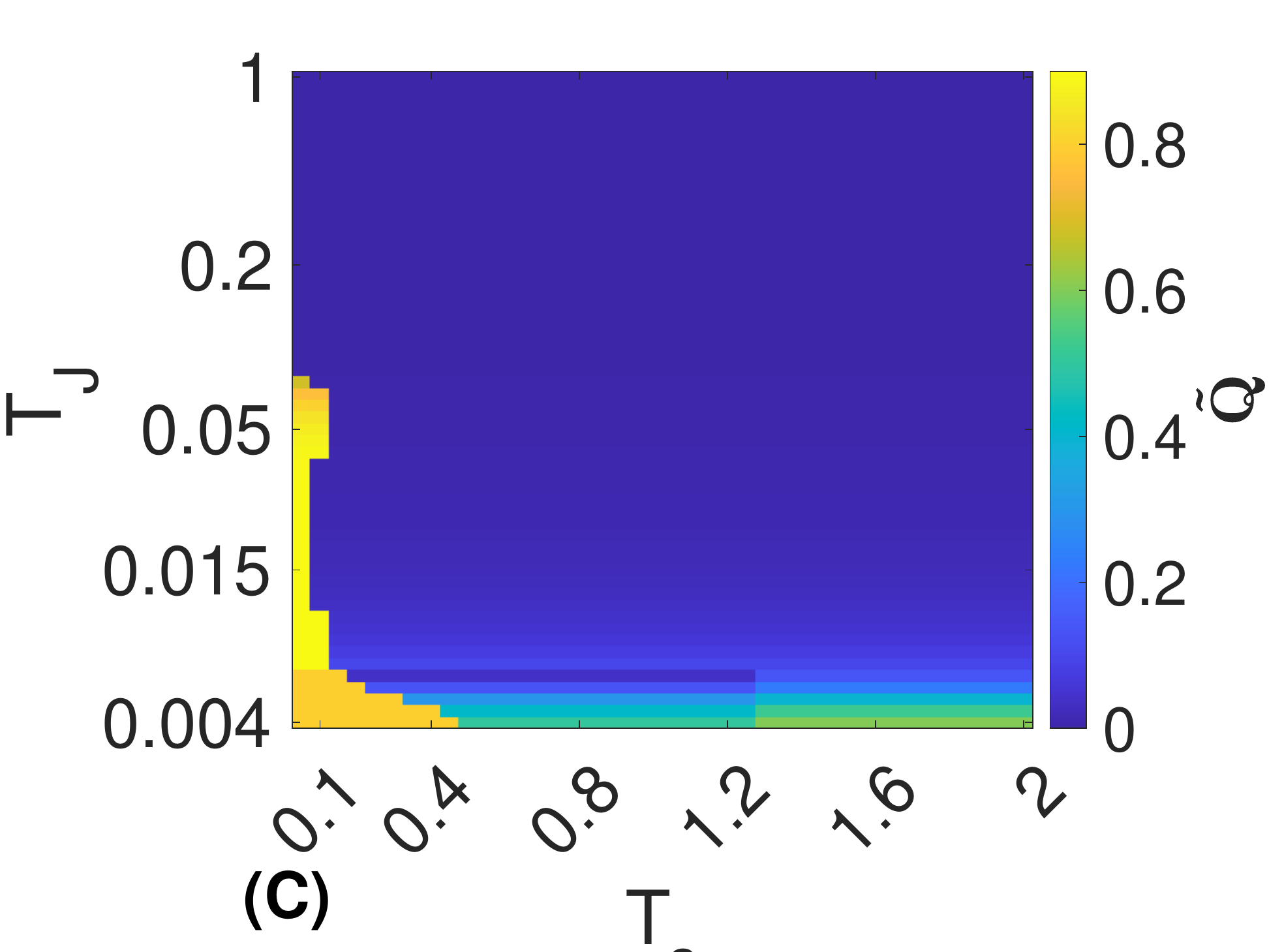}
\includegraphics[scale=0.25]{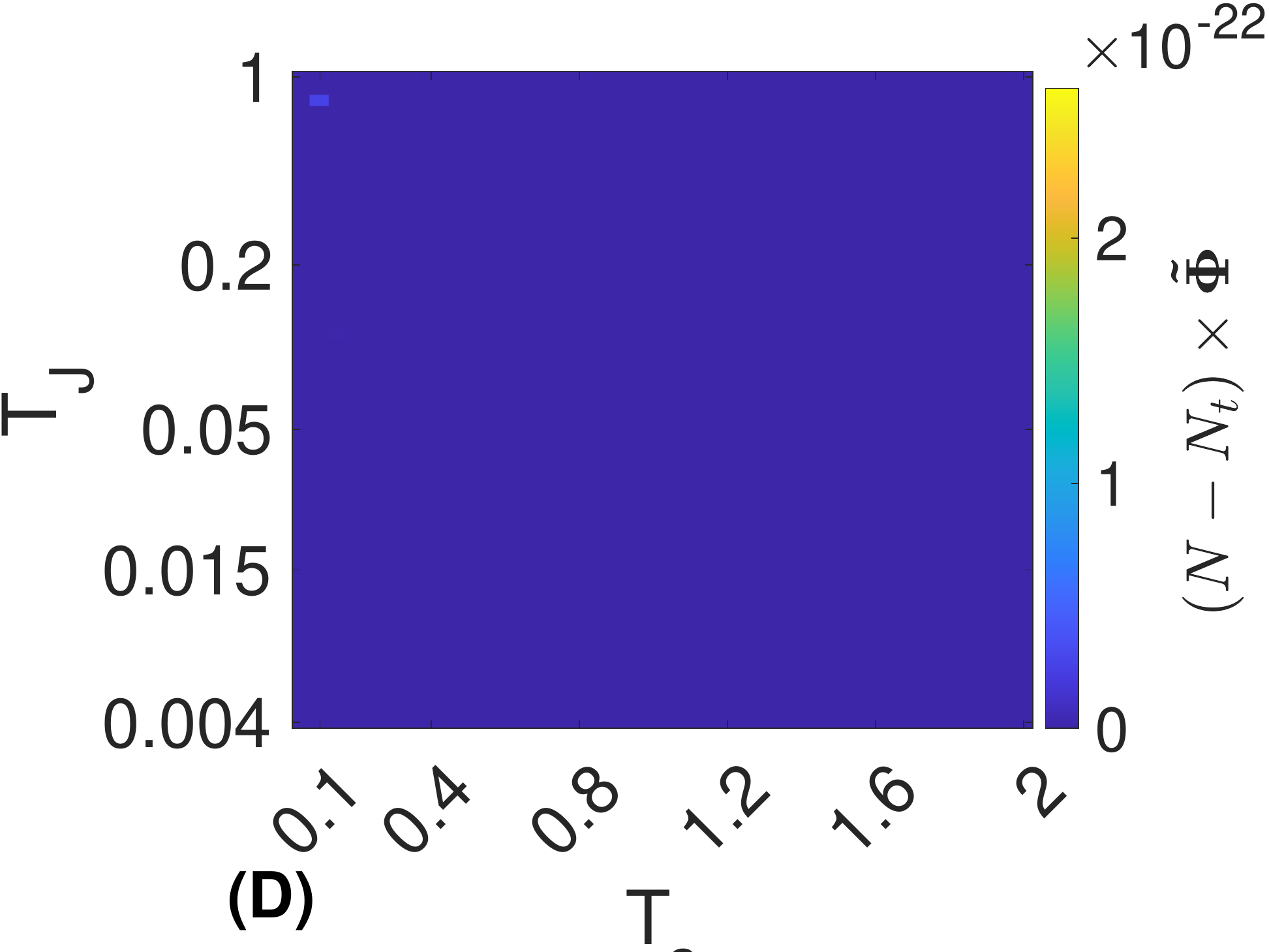}
\caption{Magnetisation  for non-target spins $\tilde{m}$ \textbf{(A)}. Overlap  between different replicas for non-target spins $\tilde{q}$ \textbf{(B)}. Averaged  correlation  of a pair of couplings between a target and a non-target spin that share a common target spin $\tilde{Q}$
\textbf{(C)}. Averaged value of the link $J^{(oo)}$ among non-target spins $\tilde{\Phi}$ \textbf{(D)}.  Here $N_t = 10$, $N = 100$. Note the $y$-axis is on logarithmic scale.} 
\label{fig:fig5}
\end{figure}

\begin{figure}[t]
    \includegraphics[scale=0.25]{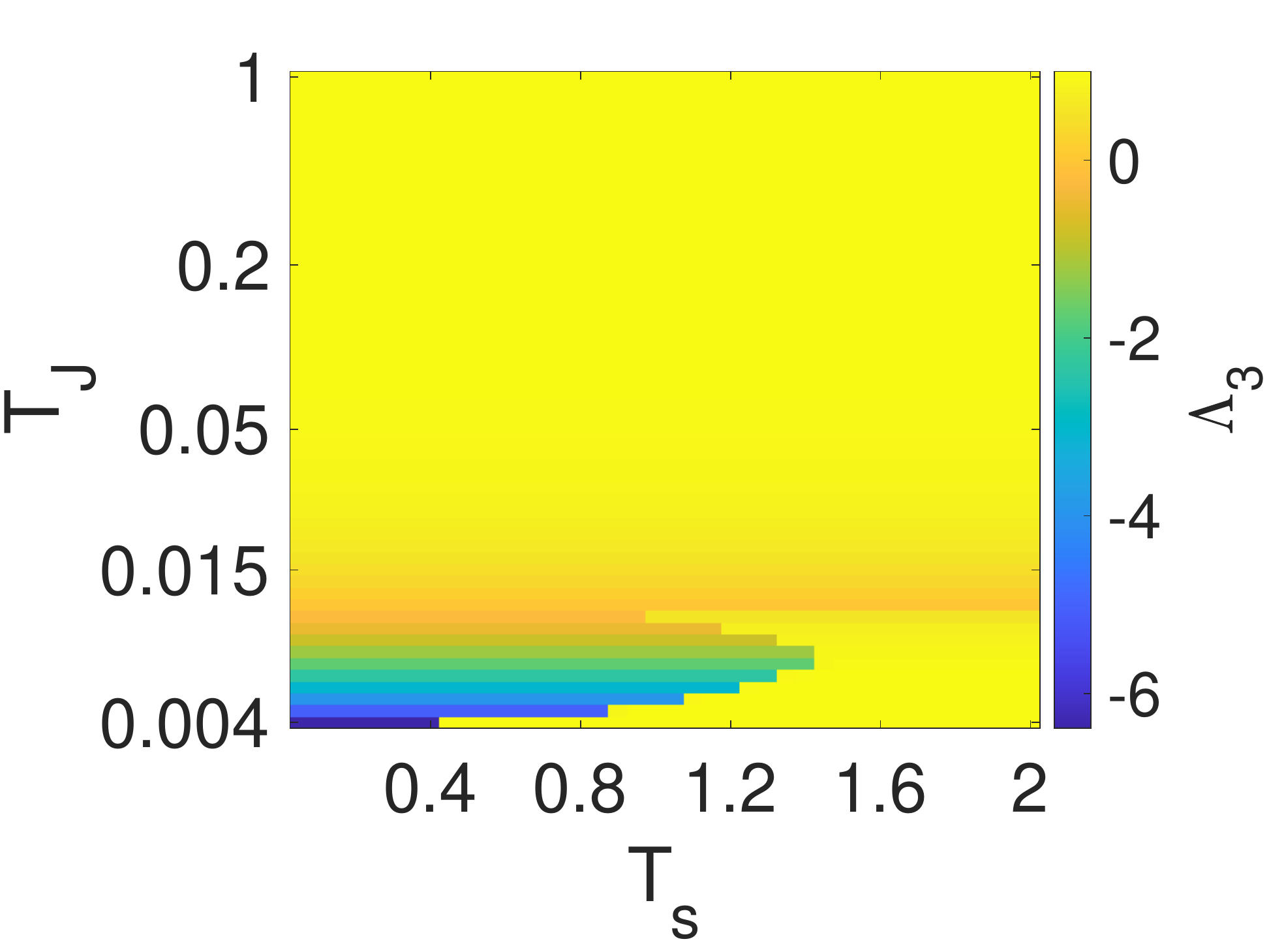}  
\caption{The third-largest eigenvalue $\Lambda_3$ of the Hessian matrix for the coupling-replicas as function of the model parameter. Here we find the region with  broken replica symmetry where  $\Lambda_3 <0$. The change in the sign of  $\Lambda_3$ happens to coincide with the transition between $Q =0$ and $Q >0$. Here $N_t = 10$ and $N = 100$. Note the $y$-axis is on logarithmic scale.}  
\label{fig:fig6}
\end{figure}

\begin{figure}[t]
         \includegraphics[scale=0.22]{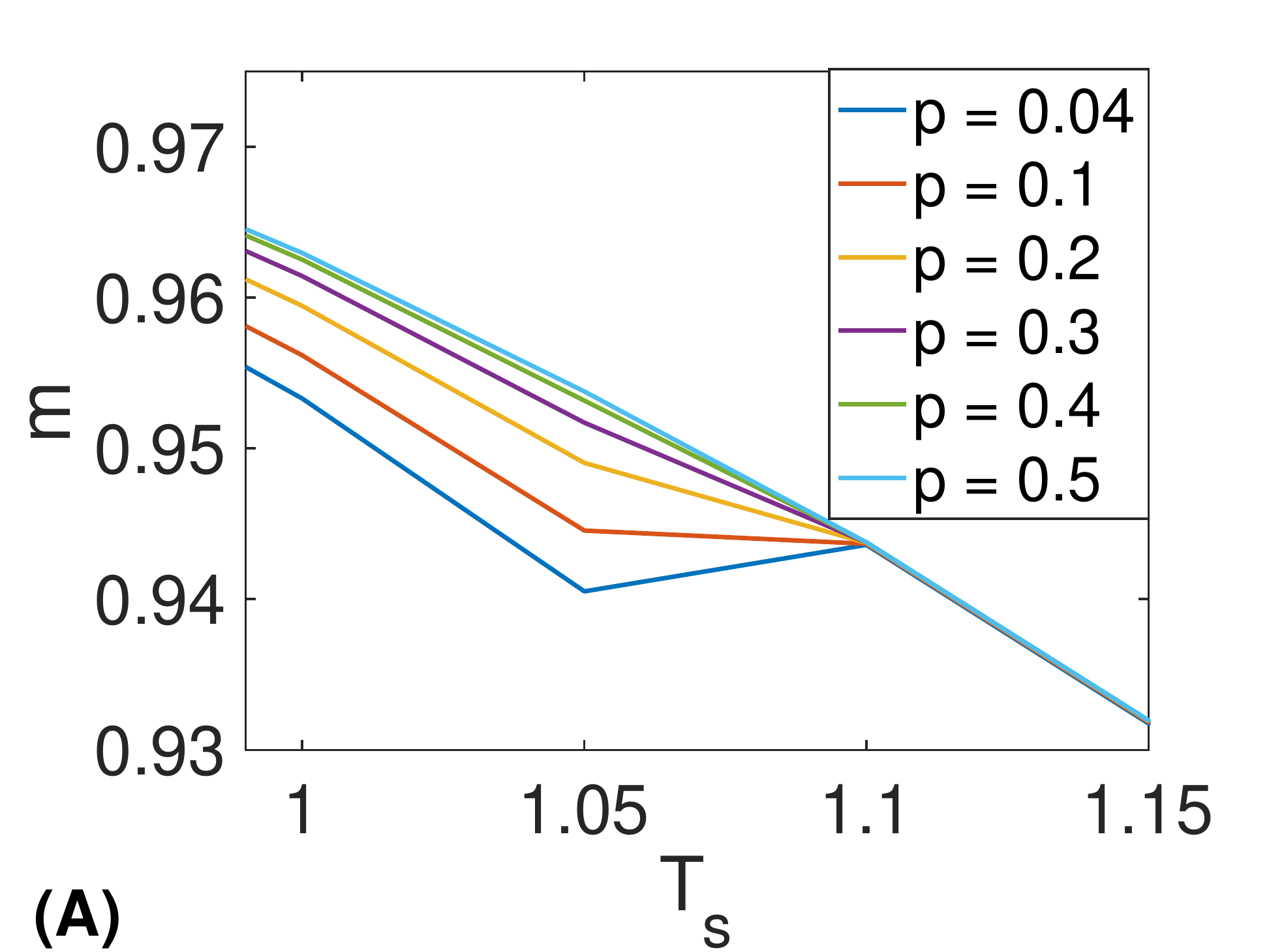}
     \includegraphics[scale=0.22]{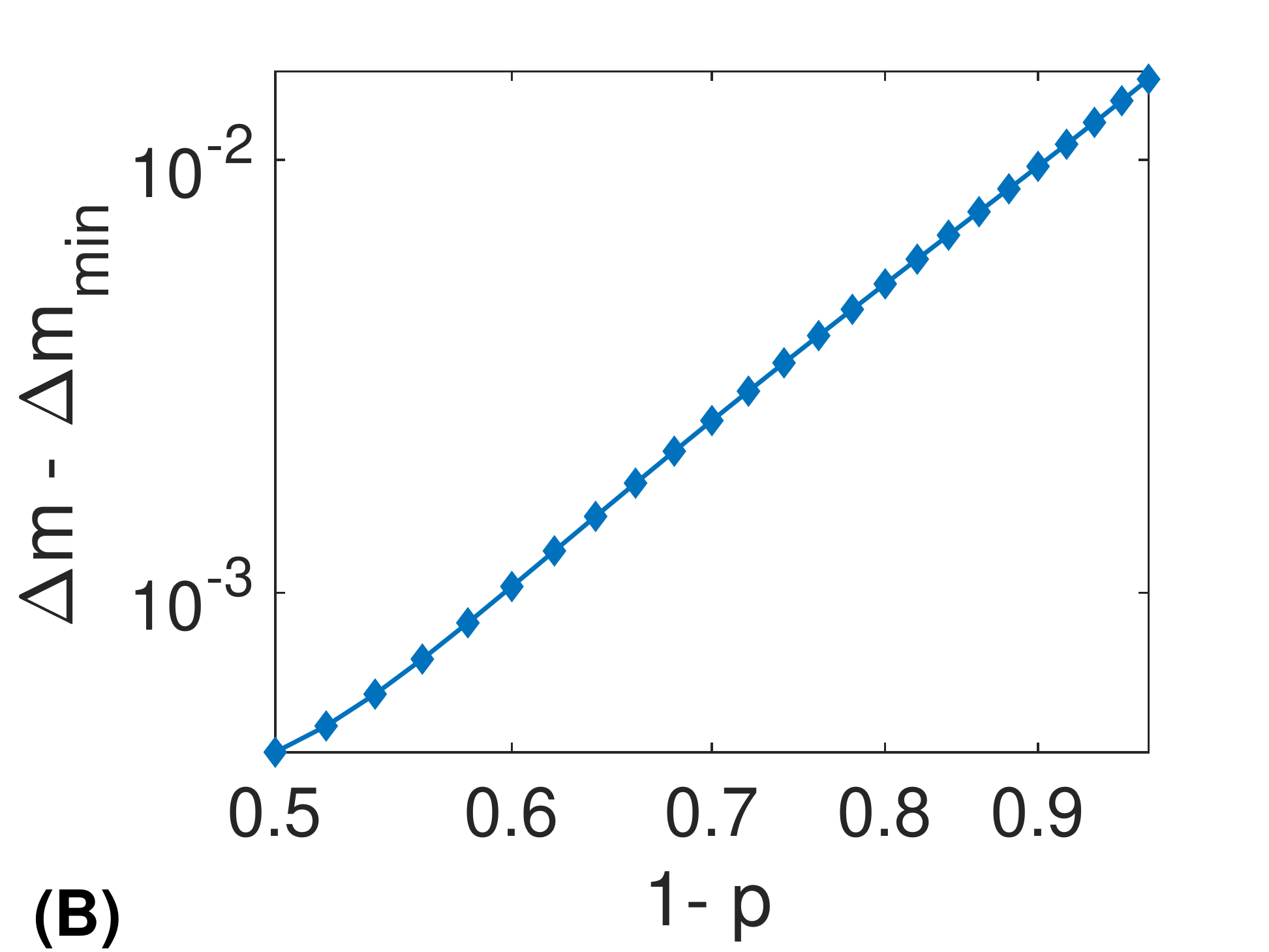}
     \includegraphics[scale=0.22]{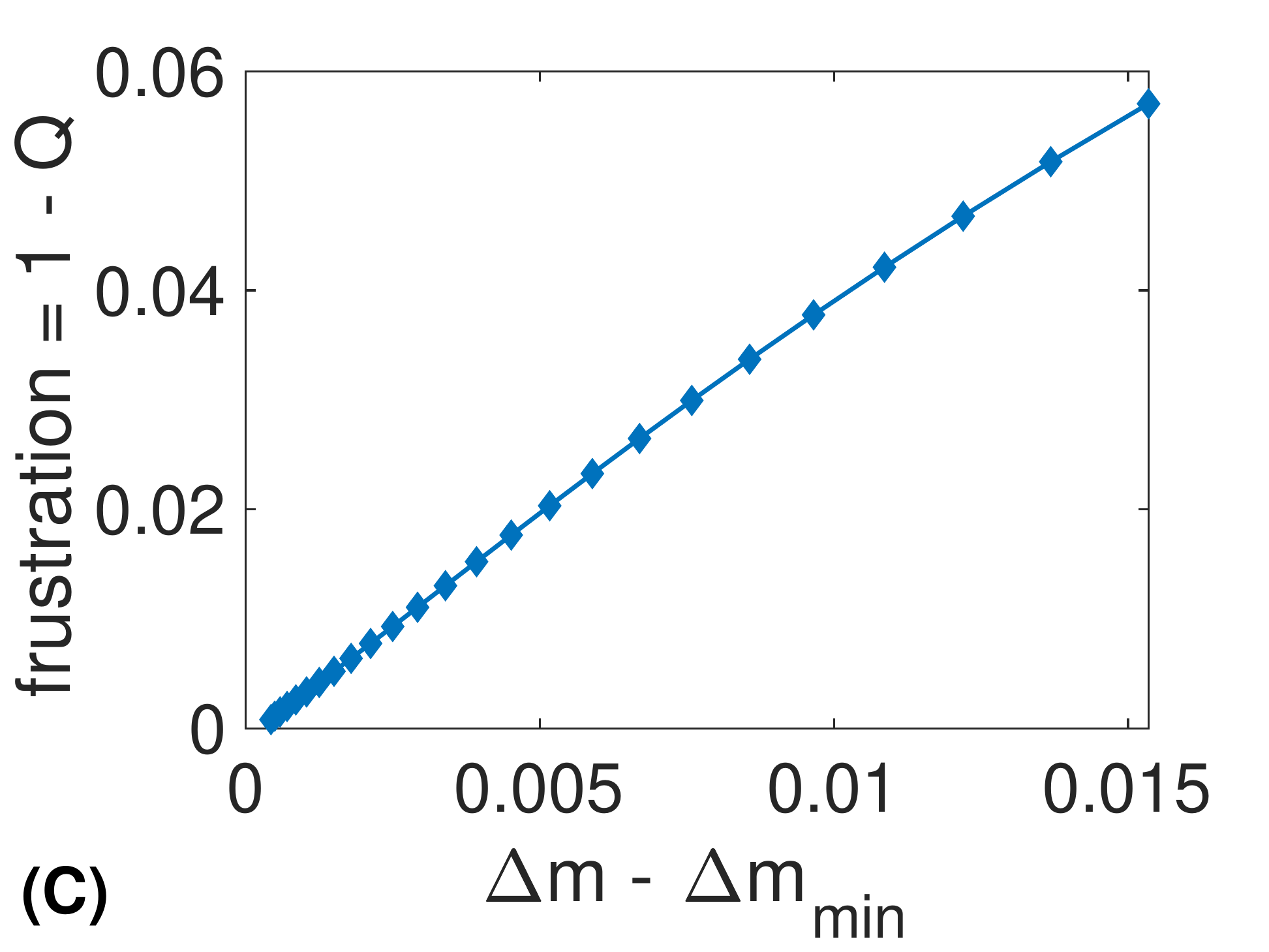}
     \includegraphics[scale=0.22]{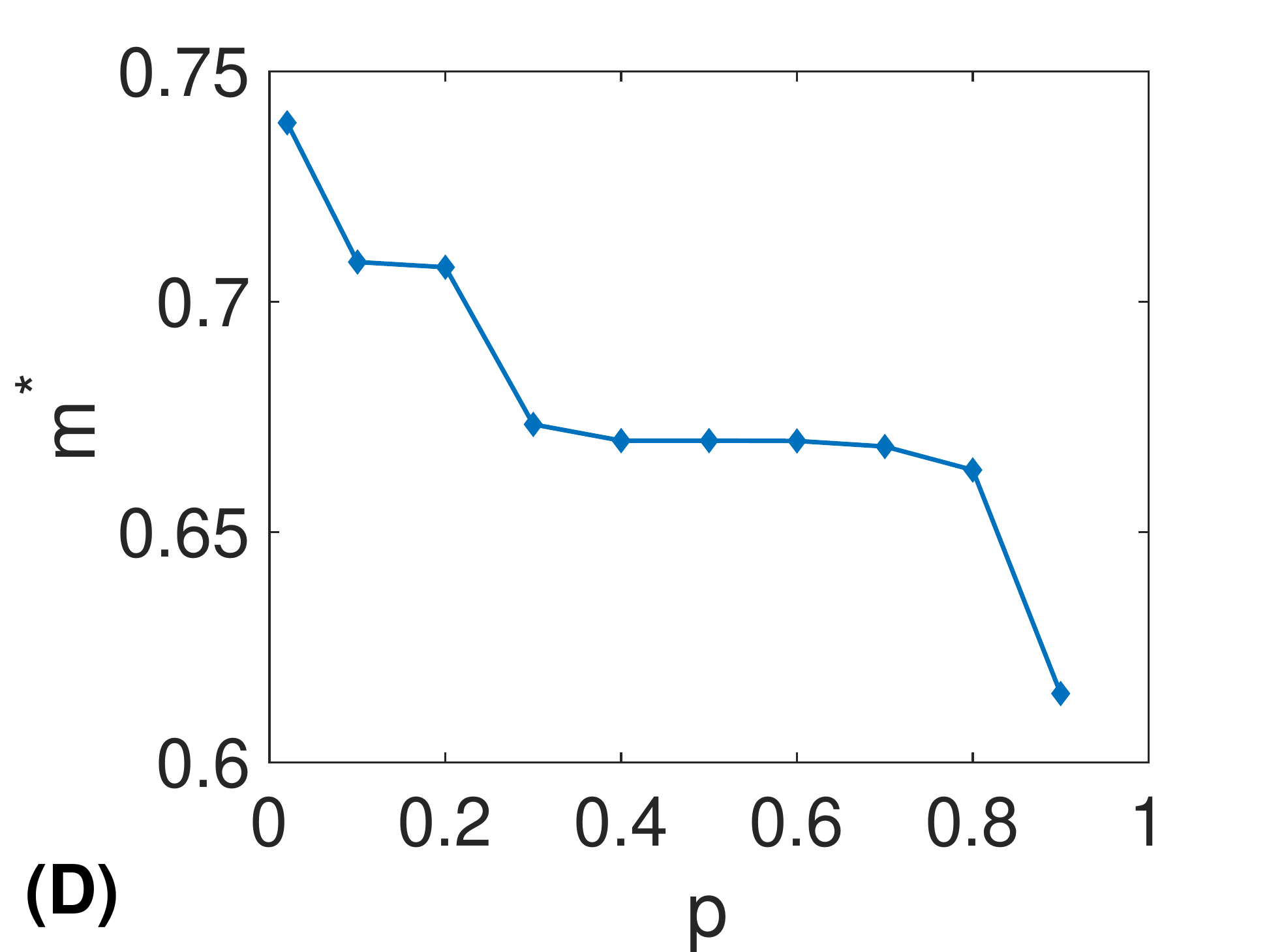}
   \caption{\textbf{(A).} A zoom-in of  Fig. \ref{fig:fig4} (A) in the region nearby $T_c^{(1)}$. \textbf{(B).} The drop of fitness subtracted from the minimal value $\Delta m_{\rm min}$ as a function of the fraction  of non-target spins. Here $\Delta m_{\rm min}$ is defined as the drop of magnetisation at the critical number of target spins $N_t^{(c)}=60$ at $T_J = 0.005$, while $\Delta m = 1 - m\big(T_c^{(1)} \big)$ is the drop of of fitness at $T_c^{(1)}$. \textbf{(C).}  Linear relationship between the frustration  $1-Q$ and $\Delta m -\Delta m_{\rm min}$ at $T_c^{(1)}$.  \textbf{(D)} The drop $m^*$ of fitness at $T_c^{(2)}$. Here $N= 100$.}
  \label{fig:fig7}
    \end{figure}

 Here we support the main text with the phase diagram for the order parameters of the non-target spins in Fig. \ref{fig:fig5} and that for the third-largest eigenvalue of the Hessian $\Lambda_3$ in Fig. \ref{fig:fig6}. We also depict the dependence of the drop in fitness at $T_c^{(1)}$ denoted by $\Delta m = 1 - m$  on the fraction of non-target spins in Fig. \ref{fig:fig7}. At the critical number of target spins $N_t^{(c)}$ this change has a minimal value $\Delta m _{\rm min}$.   Once  subtracted $\Delta m$ from  $\Delta m _{\rm min}$, we find a linear relationship exists between $\Delta m  -\Delta m _{\rm min}$ and $1-Q$.  Such  relationship is  demonstrated Fig. \ref{fig:fig7}   \textbf{(C)}. %(ii)  an increase of $Q$ from zero  to a non-zero value  is accompanied by a decrease in $m$  to a lower value. This fact suggests the existence of a trade-off relationship between $Q$ and  $m$ in the robust phase. 
 %{\re While we can not explain the origin of this trade-off within the current framework, if interpreting  $Q$ as the     genetic homogeneity,  the trade-off simply reflects the increase of   fitness with increasing redundancy of genotypes ($1-Q$). In Fig. \ref{fig:fig4}   \textbf{(C)} $1-Q$  was found to behave as an  increasing function of the number of non-target spins. %  and  seems to undergo a transition at a value $N_t^{(c)}\simeq  60$. This behavior may provide an intuitive understanding of  the trade-off as follows. Once an \emph{adapted} set of genes  that  yields relatively  high fitness is robustly established through evolution,  changes   from this state towards those of higher  fitness due to small genotypic variations  are  constrained by its genetic structure. As such  changes  are more likely to occur if there are many more  pathways along which the system can escape from the current fitness maxima. Assuming the number of such  pathways grows  with the  number of non-target spins, and combining it with the observed behaviour of $1-Q$, we arrive at the trade-off between between fitness $m$ and genetic homogeneity $Q$. 
 We finally measure how much fitness changes under a transition from robust to paramagnet phase at $T_c^{(2)}$. We denote this kind of drop by $m^*$ in Fig.  \ref{fig:fig7}   \textbf{(D)}, where we find a decrease of $m^*$ with increasing $p$, implying that fitness is more robust at higher $p$.

 \end{document}